\documentclass[review]{elsarticle}

\usepackage{lineno,hyperref}
\modulolinenumbers[5]

\usepackage[nolist]{acronym}
\usepackage{amsmath,amsthm,amssymb,epsfig,bm,amsmath}
\usepackage{array}
\usepackage{float}
\usepackage{graphicx}
\usepackage{arydshln}
\usepackage{cleveref}
\usepackage{graphicx}
\usepackage{float}
\usepackage{csquotes}
\usepackage{amsmath}
\usepackage{mhchem}
\usepackage{xparse}
\usepackage{adjustbox}
\usepackage{MnSymbol}
\usepackage{mathdots}
\usepackage{makecell}
\usepackage{listings}
\usepackage{footnote}
\usepackage[flushleft]{threeparttable}
\usepackage{multirow}
\usepackage{relsize}
\usepackage{lettrine}
\usepackage{courier}
\usepackage{comment}
\usepackage{subcaption}
\usepackage{color} 
\definecolor{mygreen}{RGB}{28,172,0} 
\definecolor{mylilas}{RGB}{170,55,241}
\usepackage{amsmath,amsthm,amssymb,amsfonts} 
\usepackage{graphicx} 
\usepackage[margin=1in]{geometry} 
\usepackage[yyyymmdd]{datetime}
\usepackage{algorithm}
\usepackage[noend]{algpseudocode}
\usepackage{lipsum}
\usepackage{setspace}
\usepackage{siunitx}
\usepackage{pdfpages}
\usepackage{tikz}
\usepackage{hhline}
\usepackage{geometry}
\usepackage{longtable}
\usetikzlibrary{positioning}
\usepackage{booktabs,caption}
\usepackage[]{hyperref}
\hypersetup{
 colorlinks  = true, 
 urlcolor   = blue, 
 linkcolor  = black, 
 citecolor  = blue 
}
\usepackage[utf8]{inputenc}
 
\usepackage{listings}
\usepackage{color}
\definecolor{codegreen}{rgb}{0,0.6,0}
\definecolor{codegray}{rgb}{0.5,0.5,0.5}
\definecolor{codepurple}{rgb}{0.58,0,0.82}
\definecolor{backcolour}{rgb}{0.95,0.95,0.92}
\lstdefinestyle{mystyle}{
  backgroundcolor=\color{backcolour},  
  commentstyle=\color{codegreen},
  keywordstyle=\color{magenta},
  numberstyle=\tiny\color{codegray},
  stringstyle=\color{codepurple},
  basicstyle=\footnotesize,
  breakatwhitespace=false,     
  breaklines=true,         
  captionpos=b,          
  keepspaces=true,         
  numbers=left,          
  numbersep=5pt,         
  showspaces=false,        
  showstringspaces=false,
  showtabs=false,         
  tabsize=2,
  escapeinside={<@}{@>},
}
 
\usepackage{booktabs} 
\usepackage{siunitx} 

\usepackage{mathtools} 
\usepackage{gensymb} 
\usepackage{mhchem} 
\usepackage{fixltx2e} 
\usepackage{bbm}

\usepackage{multirow}

\usepackage{fancyhdr} 
\theoremstyle{definition}

\theoremstyle{definition}

\theoremstyle{remark}

\usepackage{soul} 
\soulregister\cite7
\soulregister\ref7
\soulregister\pageref7
\usepackage{bm}

\usepackage{framed} 

\usepackage{multicol} 

\usepackage{nomencl} 
\usepackage{tikz}
\makenomenclature
\usepackage[resetlabels,labeled]{multibib}

\renewcommand*\nompreamble{\begin{multicols}{2}}

\renewcommand*\nompostamble{\end{multicols}}

\usepackage{amssymb}
\usepackage{pifont}
\definecolor{light-gray}{gray}{0.95}

\journal{Journal Review}

\begin{document}

\begin{frontmatter}

\title{\large On the Impact of Monte Carlo Statistical Uncertainty on Surrogate-based Design Optimization}

\author{Omer F. Erdem$^{a,*}$, David P. Broughton$^{b}$, Josef Svoboda$^{b}$, Chengkun Huang$^{b}$, Majdi I. Radaideh$^{a,*}$}

\cortext[mycorrespondingauthor]{Corresponding Author: M. I. Radaideh (radaideh@umich.edu), Omer F. Erdem (oferdem@umich.edu)}

\address{$^{a}$Department of Nuclear Engineering and Radiological Sciences, University of Michigan, Ann Arbor, MI 48109, United States}
\address{$^{b}$Los Alamos National Laboratory, Los Alamos, NM 87545, United States}

\begin{abstract}

\small

In multi-objective design tasks, the computational cost increases rapidly when high-fidelity simulations are used to evaluate objective functions. Surrogate models help mitigate this cost by approximating the simulation output, simplifying the design process. However, under high uncertainty, surrogate models trained on noisy data can produce inaccurate predictions, as their performance depends heavily on the quality of training data. This study investigates the impact of data uncertainty on two multi-objective design problems modelled using Monte Carlo transport simulations: a neutron moderator and an ion-to-neutron converter. For each, a grid search was performed using five different tally uncertainty levels to generate training data for neural network surrogate models. These models were then optimized using NSGA-III. The recovered Pareto-fronts were analyzed across uncertainty levels, and the impact of training data quality on optimization outcomes was quantified. Average simulation times were also compared to evaluate the trade-off between accuracy and computational cost. Results show that the influence of simulation uncertainty is strongly problem-dependent. In the neutron moderator case, higher uncertainties led to exaggerated objective sensitivities and distorted Pareto-fronts, reducing normalized hypervolume. In contrast, the ion-to-neutron converter task was less affected—low-fidelity simulations produced results similar to those from high-fidelity data. These findings suggest that a fixed-fidelity approach is not optimal. Surrogate models can still recover the Pareto-front under noisy conditions, and multi-fidelity studies can help identify the appropriate uncertainty level for each problem, enabling better trade-offs between computational efficiency and optimization accuracy.

\end{abstract}


\end{frontmatter}

\section{Introduction}
\label{sec:intro}


High-fidelity simulations play a critical role in advancing many different engineering disciplines by providing accurate and detailed insights into radiation transport \cite{Miller2010}, complex fluid dynamics \cite{cfd_best_practices}, structural responses \cite{radaideh2022bayesian}, thermal behavior \cite{wang2015towards}, infrastructure design \cite{wang2010high}, electronics design \cite{singh2024rapid}, and safety analysis \cite{radaideh2018criticality}, among many other phenomena. In nuclear engineering, high-fidelity Monte Carlo neutron transport codes—such as the MCNP code~\cite{MCNP63}, GEANT \cite{geant4}, FLUKA \cite{bohlen2014_fluka}, PHITS \cite{sato2018_phits}, Serpent \cite{serpent2}, and many other—provide precise simulations of neutron interactions \cite{sharma2014mcnp,tekin2017simulations,price2022multiobjective, zavorka2018markIVdesign, broughton2025commissioning} and criticality calculations \cite{kromar2013comparison,price2019advanced}, thereby enabling the design and optimization of radiation transport systems with robust safety margins. These simulations are indispensable for neutron generator optimization, moderator–reflector design, fuel and accelerator studies, and non-proliferation analysis, as experimental validation is often prohibitively expensive or infeasible \cite{Miller2010,Babcock2008,Wilson2011,Martinez2012}. However, the high computational cost of these simulations remains a major challenge, often requiring high-performance computing (HPC) resources and parallelized algorithms to achieve feasible runtimes \cite{hpc_reactor_physics}. 

Designing complex engineered systems often involves expensive, uncertainty-prone simulations, posing challenges for ensuring system reliability and safety. This study addresses a critical gap at the intersection of system safety, performance, and design optimization under uncertainty by investigating how training data uncertainty—originating from Monte Carlo particle transport simulations—affects surrogate-assisted multi-objective optimization. By systematically quantifying the impact of simulation-induced tally uncertainties on surrogate model fidelity and optimization outcomes, this work contributes to a deeper understanding of model-driven decision-making under uncertainty. The insights derived not only enhance the reliability of surrogate-based design tools but also inform the development of robust optimization strategies for high-dimensional, safety-critical systems.

Recent research has significantly advanced the use of Monte Carlo simulations in design optimization under uncertainty, particularly within engineering and reliability systems. Several studies integrated Monte Carlo methods with other computational techniques to handle both aleatory and epistemic uncertainties. Miska and Balzani \cite{miska2025reliability} proposed a reliability-based design optimization framework using extended Optimal Uncertainty Quantification, while Jia et al. \cite{jia2025hierarchical} applied hierarchical Bayesian modeling for improved reliability assessment using static and dynamic data. Zhang and Zhao introduced linear moments to model unknown distributions in reliability analysis \cite{zhang2024linear}, and Thaler et al. combined subset simulation with Hamiltonian neural networks for efficient reliability computations \cite{thaler2024reliability}. In energy systems, Monte Carlo simulations were integrated with genetic algorithms to optimize hybrid designs under variable conditions \cite{farh2024optimization}.

Recent advancements in reliability engineering and uncertainty quantification have led to the development of sophisticated modeling, simulation, and optimization techniques across complex systems such as structural components, nuclear power plants, and multidisciplinary designs. For instance, Jung et al. proposed a framework that integrates aleatory and epistemic uncertainties into statistical model calibration and design optimization to improve the robustness of engineering systems \cite{jung2022statistical}. In the context of real-time emergency response, Cui et al., \cite{cui2025timeliness} developed a rush repair scheduling model that minimizes workforce response time during facility emergencies, emphasizing operational timeliness under uncertainty. Similarly, the study by \cite{nannapaneni2020probability} introduced a probability-space surrogate modeling approach for efficient multidisciplinary optimization under uncertainty, reducing computational costs while maintaining accuracy.

Addressing structural performance, Zhao et al. \cite{zhao2025uncertainty} proposed a dynamic topology optimization method for cross-scale structures that accounts for uncertainty during the design process. In nuclear applications, Xiao et al. \cite{xiao2025reliable} developed a reliable and adaptive prediction framework for nuclear power plant systems using advanced learning algorithms to support proactive maintenance. Zeng et al. \cite{zeng2025development} explored dynamic simulations with uncertainty analysis to assess the behavior of engineering systems with variable parameters, Radaideh et al. used Bayesian model averaging for uncertainty propagation in nuclear computer codes \cite{radaideh2019integrated}, and Chen et al. \cite{chen2025dynamic} assessed the dynamic performance of coordinated control systems in small modular reactors under parameter uncertainty.

\subsection{Technical Challenges in Nuclear Design Optimization}

Nuclear science and engineering domains encompass a wide range of complex physical phenomena, including neutron transport, radiation shielding, reactor physics, and particle interactions in high-energy systems. One of the fundamental computational challenges in optimization in nuclear science and engineering arises from the stochastic nature of particle interactions. Monte Carlo neutron transport codes, such as the MCNP code, rely on probabilistic sampling of individual particle histories to estimate physical quantities like neutron flux, reaction rates, and dose distributions \cite{MCNP63}. However, the effectiveness of these simulations is hindered in certain scenarios due to inherently low interaction probabilities (depending on reaction cross-sections). For instance, in neutron moderators, the fast neutrons undergo multiple scattering events before being moderated. The probability of a neutron interacting within a specific region is relatively low, leading to high statistical uncertainty in simulation results. The accuracy of these simulations is limited in certain situations due to the very low chances of particle interactions (which depend on nuclear reaction probabilities), particularly in scenarios involving high-energy neutron sources where interaction cross sections are small. In cases like optimizing a moderator-reflector setup for a 4$\pi$ isotropic 14 MeV neutron source within a constrained geometry capturing only ~3 sr (steradian), Monte Carlo simulations become highly inefficient. Similarly, in proton converters used for medical and accelerator applications, the probability of protons generating secondary particles, such as neutrons through spallation reactions, is typically low due to the low reaction cross-section, requiring a vast number of simulated histories to achieve statistically meaningful results \cite{nakayama2021jendl,kunieda2016jendl}. 

These challenges necessitate running MCNP simulations with a high number of particle histories—referred to as NPS (Number of Particle Simulations)—to reduce variance and obtain reliable statistical estimations. In an optimization routine, these simulations must be executed thousands of times to find optimal design configurations, significantly increasing computational costs. Due to the low interaction probabilities and the large number of required simulation runs, the first step should involve improving simulation efficiency through variance reduction techniques (e.g., source biasing, weight windows), followed by automation and optimization processes to refine the system design. After this step, high-performance computing (HPC) resources and surrogate models are employed to mitigate computational costs. These methods enhance simulation efficiency, making large-scale Monte Carlo simulations feasible within time and statistical accuracy constraints, but the challenges of repetitive executions aggravate HPC needs to solve optimization tasks. This has led various development teams to apply optimization techniques in different ways. One tested option for physical design optimization is DAKOTA \cite{adams2020dakota}, developed by Sandia National Laboratories. It can be integrated with MCNP geometry generators and supports simulations using Unstructured Mesh (UM) in MCNP, as demonstrated in the design of the Second Target Station at Oak Ridge National Laboratory \cite{ZAVORKA2023_Hybrid, GHOOS2024_optimization}. While powerful optimization tools already exist, there remains a strong interest in developing new tools with modernized approaches. Such software is especially valued for its ease of distribution and lower cost, making it well-suited for the daily optimization of smaller-scale projects.

Given the high computational cost of high-fidelity nuclear simulations, various acceleration techniques have been developed to enhance efficiency while preserving accuracy. Surrogate modeling, including machine learning-based and physics-informed approaches, has emerged as a powerful tool to approximate complex simulations at a fraction of the computational cost \cite{surrogate_models_nuclear}. The integration of artificial intelligence (AI) and machine learning (ML) in surrogate modeling has shown promise in accelerating these high-fidelity simulations without compromising reliability \cite{surrogate_models_progress, dutta2020surrogate, palar2019surrogate}. Deep neural network surrogates \cite{radaideh2019combining,saleem2020application,radaideh2020analyzing} and Gaussian process regression \cite{radaideh2020surrogate} approaches have demonstrated the potential to drastically reduce simulation times—by orders of magnitude in some cases in nuclear science and engineering applications. In addition, surrogate models preserved the essential physical accuracy required for critical radiation transport analyses, as demonstrated in these studies \cite{Smith2019,DoeRoe2020,Lee2021}. Alternative to surrogate modeling of the problems, multi-fidelity modeling leverages a combination of low-fidelity and high-fidelity simulations to balance computational efficiency and predictive accuracy, enabling more effective uncertainty quantification in reactor analysis \cite{multifidelity_methods}. Adaptive sampling techniques strategically select simulation points to optimize computational resources, reducing the number of expensive evaluations required for convergence \cite{adaptive_sampling}. Bayesian optimization has been successfully applied to optimize reactor core parameters and control strategies with minimal simulation runs \cite{bayesian_optimization}. In addition, parallel computing techniques, such as multi-threading and GPU acceleration, significantly reduce runtime by distributing workloads across multiple cores or exploiting massively parallel architectures, making it feasible to conduct Monte Carlo simulations more efficiently \cite{gpu_monte_carlo}. The integration of these methods is crucial for accelerating design, optimization and safety assessments particularly in the fields that require extensive computational studies to meet regulatory and economic constraints.

One of these acceleration techniques is using surrogate models to imitate the original problem. Surrogate models are simplified representations that approximate the behavior of complex systems, enabling faster simulations and analyses. In nuclear science problems, surrogate models are invaluable for tasks like uncertainty quantification, optimization, and real-time decision-making, where high-fidelity models may be computationally prohibitive. Various types of surrogate models are employed, each with distinct strengths and weaknesses. Polynomial response surface models provide explicit mathematical relationships between inputs and outputs, making them easy to interpret; however, they struggle with capturing highly nonlinear behaviors \cite{chen2024surrogate}. Gaussian Process Regression (Kriging) offers high accuracy and built-in uncertainty quantification but can be computationally intensive, especially with large datasets \cite{figueroa2020gaussian}. Artificial Neural Networks (ANNs) are capable of modeling complex nonlinear relationships and can detect all possible interactions between predictor variables; however, they are often criticized for their "black box" nature, greater computational burden, and proneness to overfitting \cite{gardner1998artificial}. Polynomial Chaos Expansions (PCE) enable analytical uncertainty propagation \cite{radaideh2022bayesian} but become impractical in high-dimensional spaces due to the rapid growth in polynomial terms. Gu et al. \cite{gu2025reliability} developed a strategy space reduction method to determine a more suitable value range for each decision variable, thereby streamlining the optimization process. The Genetic Algorithm is then utilized to find the optimal solution within this refined strategy space. Jiang et al. \cite{jiang2025development} introduced a CNN-based surrogate model for structural health monitoring, facilitating early detection of anomalies and supporting maintenance decisions of damage of pipelines buried in spatially varied soils. Finally, this study \cite{keramatinejad2025hybrid} proposed a hybrid framework combining adaptive surrogate modeling and sampling methods to enhance reliability assessment in multidisciplinary design optimization tasks. The choice of surrogate model depends on the specific application, with trade-offs between accuracy, interpretability, data needs, and computational efficiency shaping their implementation.

While surrogate models play a crucial role in reducing computational costs for complex simulations, they inherently introduce uncertainties that must be quantified to ensure reliable predictions, especially when coupled with optimization algorithms like genetic algorithms. Various approaches have been developed to address the uncertainties associated with surrogate modeling, ranging from Bayesian methods, Gaussian process regression, and ensemble methods to physics-constrained learning techniques \cite{Zhu2019}. For instance, reliability-based design optimization (RBDO) integrates uncertainty quantification techniques such as equivalent reliability indices to assess the impact of surrogate model uncertainty on system design \cite{Li2019}. Furthermore, data-driven surrogate modeling approaches, such as those utilizing closed observables and diffusion maps, have demonstrated significant improvements in real-time uncertainty quantification \cite{Dietrich2018}. \textit{While these methods effectively quantify the interpolation uncertainty of the surrogate model, they do not directly account for the uncertainty in the training data itself, which originates from the Monte Carlo simulation and propagates into the surrogate model parameters}.

Design optimization under uncertainty involves identifying the best possible design solutions while considering variability and uncertainty in system parameters, operating conditions, or surrogate models used to approximate physical phenomena. This type of optimization is widely applied across various fields \cite{sahinidis2004optimization,radaideh2020design,jaeger2013aircraft}. Surrogate-assisted optimization has gained popularity as a means to reduce the computational cost associated with repeatedly executing high-fidelity solvers by leveraging surrogate models, as demonstrated in multiple studies \cite{radaideh2022model,price2022multiobjective}. However, ensuring the accuracy of optimization results requires an equally accurate surrogate model, which demands the generation of large datasets—an impractical challenge when using Monte Carlo codes that inherently carry uncertainty. This challenge becomes increasingly complex as the number of objectives to optimize grows, particularly in multi-objective problems \cite{erdem2025multi}. Several optimization techniques can be employed with surrogate models for engineering applications. These methods include gradient-based algorithms \cite{peri2012multistart,jiaqi2022gradient}, metaheuristic algorithms \cite{fathnejat2020efficient,price2022multiobjective}, multi-arm bandit and reinforcement learning approaches \cite{slivkins2019introduction,radaideh2021physics,radaideh2021rule}, as well as hybrid ensemble techniques \cite{lynn2017ensemble,price2023animorphic,radaideh2022pesa,radaideh2021large}. \textit{A significant research gap remains in quantifying the impact of Monte Carlo sampling uncertainty on the performance of optimization routines when these solvers are used to train surrogate models}. For instance, while the study by \cite{gu2023openneomc} successfully integrates optimization routines with the Monte Carlo code OpenMC for criticality calculations, the authors had to reduce the NPS to optimize computational cost, which indirectly affects the quality of the optimization solution.

\subsection{Novelty}

This research explores surrogate-assisted optimization in expensive Monte Carlo particle transport simulations by systematically evaluating the impact of training data uncertainty caused by Monte Carlo sampling on multi-objective design problems—an area that has not been thoroughly explored in the existing literature. Utilizing Monte Carlo neutron transport simulations with varying tally uncertainties, we generated low-fidelity and high-fidelity datasets. For both problems, a grid search of the design space was performed with five different tally uncertainty thresholds—where tally refers to a user-defined measurement, such as neutron flux, within the simulation. Using the created low-fidelity and high-fidelity datasets, we trained feedforward neural network (FNN) surrogate models for two critical design tasks: neutron moderator optimization and ion-to-neutron converter design. By conducting a controlled grid search with five different levels of tally uncertainty for each problem, we establish a structured methodology to quantify the effect of training data uncertainty on surrogate model accuracy and optimization performance. Furthermore, we employ Non-dominated Sorting Genetic Algorithm III (NSGA-III) multi-objective optimization to analyze how uncertainty affects the recovery of the Pareto-front, challenging the assumption that surrogate models remain effective across all uncertainty levels. Our findings provide a novel perspective on the relationship between training data accuracy and optimization success, offering valuable insights into the feasibility of surrogate-assisted multi-objective optimization under realistic uncertainty constraints.  This study establishes guidelines for selecting appropriate training data quality when deploying FNN-based surrogate models in high-cost, high-dimensional design tasks. Accordingly, the key contributions and findings of this study can be summarized as follows:  

\begin{enumerate}  
    \item This study provides one of the first detailed evaluations of how varying Monte Carlo tally uncertainties affect surrogate model accuracy and the ability to recover accurate Pareto fronts.

    \item A reliability analysis of neural network surrogates trained on uncertain data is conducted, offering practical guidance for selecting appropriate training data accuracy to ensure reliable optimization results.

    \item The results show that the effect of simulation uncertainty on optimization is problem-specific. In some cases, high-fidelity data is essential, while in others, surrogate-driven optimization can recover the Pareto-front by using lower-fidelity data.

    \item It is demonstrated that low-fidelity simulations can sometimes distort predictions, reducing the quality of the Pareto-front. However, in other problems, they can yield comparable optimization outcomes while significantly reducing computational cost.
\end{enumerate} 

The remaining sections are organized as follows: Section \ref{sec:problems} explains two selected neutron transport problems and details of the data collection process in these problems. This section explains the Monte Carlo neutron transport simulations and uncertainty variations used for training the FNN surrogate models. In Section \ref{sec:method}, we outline the research methodology, covering the surrogate modeling approach, NSGA-III optimization framework, and the evaluation of Pareto-front recovery under different uncertainty conditions. Following this, Section \ref{sec:results} presents the results and discussions, interpreting the impact of uncertainty on optimization outcomes. Finally, Section \ref{sec:conc} concludes the paper, summarizing key findings and suggesting directions for future research.

\section{Problem Definitions}
\label{sec:problems}



Two separate problems are investigated in this research: the neutron moderator design problem and the ion-to-neutron converter design problem. The problems mentioned here are simulated with well-established Monte Carlo radiation transport code the MCNP code \cite{MCNP63}. The geometry, source, and tally definitions of these selected problems are explained in the following subsections.

\subsection{Neutron Moderator Problem}

\begin{figure}[!h]
    \centering
    \includegraphics[width=0.5\textwidth]{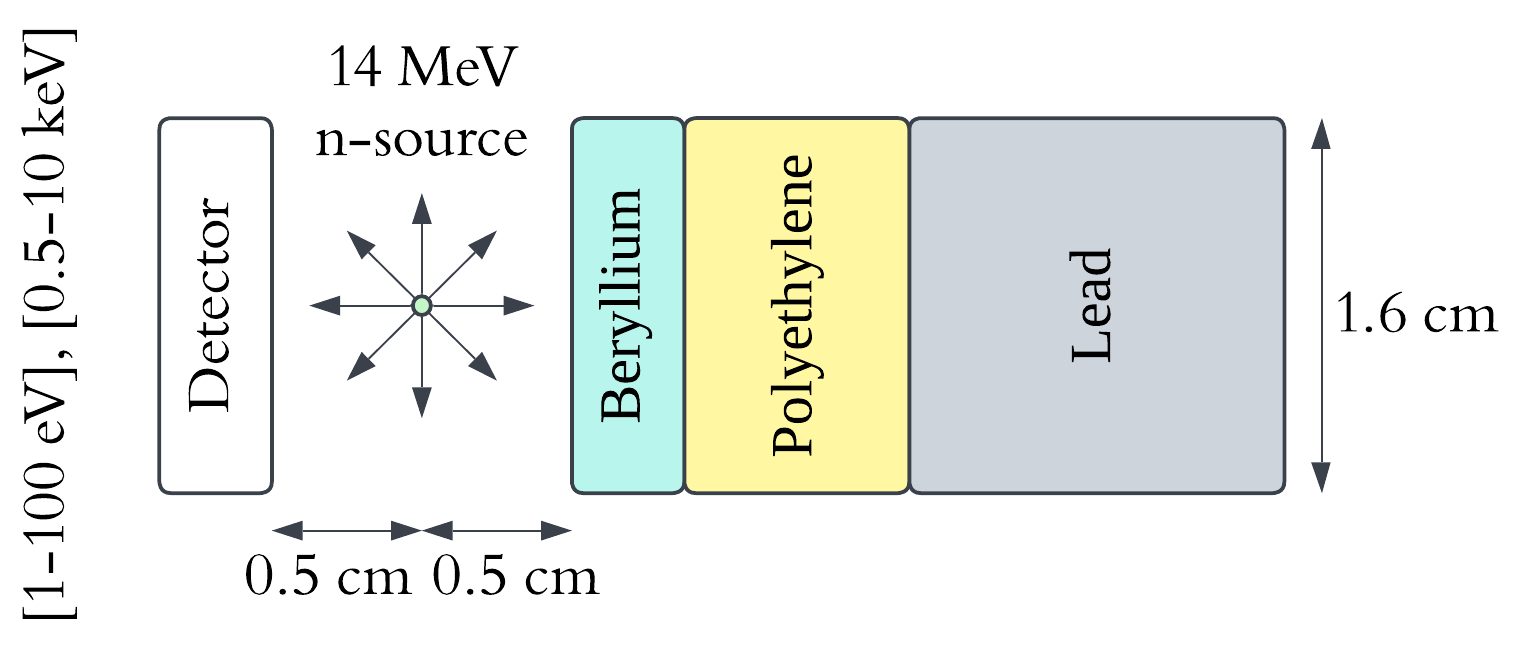}
    \caption{Geometry definition of the moderator design problem. The Be, PE and Pb layer dimensions are parametric in the problem, and are not drawn to size in this figure. Detector and moderator regions have cylindrical shapes.}
    \label{fig:refl_prob_geom}
\end{figure}

The goal of this design problem is to enhance the flux of lower-energy neutrons, in the ranges of [1–100~eV] and [0.5–10~keV], moderated from a 14~MeV, $4\pi$ isotropic neutron source, with emphasis on the neutrons directed along the normal to the moderator–reflector surface. To investigate this, a simplified moderator geometry is constructed under the constraint of capturing a solid angle of approximately $3~sr$. This geometry consists of three concentric cylindrical regions composed of beryllium (Be), polyethylene (PE), and lead (Pb). The configuration of these materials is illustrated in Figure~\ref{fig:refl_prob_geom}.

This moderator–reflector assembly is designed to optimize the back-scattered energy spectrum of neutrons by combining moderation and reflection. In this configuration, the Be layer serves as an efficient converter, using its inelastic scattering properties to down-scatter the fast 14~MeV neutrons from the source into a broader energy range. Although the thin Be layer exhibits minimal direct interaction with the initial fast neutrons, it helps to converting these neutrons into energies suitable for further moderation. With and without Be geometries were studied and Be adds up to {4.26\% of the low energy neutron yield} without further moderating the higher energy neutrons. The polyethylene (PE) layer, owing to its high hydrogen content, further moderates the fast neutrons through successive elastic collisions, thereby efficiently reducing their energy. Finally, the lead (Pb) layer—characterized by its high density and atomic number—acts as a robust reflector, scattering both fast and moderated neutrons back toward the PE moderator and the Be converter, thus enhancing overall back-scattering efficiency. This synergistic combination ensures that the incident high-energy neutrons are converted into a narrowly defined energy spectrum. \textit{The thickness values of the moderator–reflector regions are selected as the design parameters to be optimized for this problem.}

In the MCNP simulations for the neutron moderator problem, the neutron source is modeled as an isotropic emitter. The D-T neutron generator is represented with a constant particle energy of 14.1\,MeV, thereby replicating the actual neutron spectrum produced in D-T fusion reactions without introducing uncertainties. The source is positioned 0.5\,cm from both the beryllium surface and the detector surface. To maximize the back-scattered neutron flux of lower-energy neutrons reaching the detector, the source-to-setup distance is minimized within the available geometrical constraints. Furthermore, the source is located along the negative x-axis relative to the Be-CH2-Pb geometry configuration, as illustrated in Figure~\ref{fig:refl_prob_geom}.

The tallies defined in this problem are circular surfaces aligned along the negative x-axis relative to the source. The tallies have a radius of 0.8 cm, and they are positioned 0.5 cm away from the source, mirroring the setup geometry. Both tallies are defined at the same position and with the same size. The first tally records the neutron flux for the 1~eV to 100~eV energy interval, while the second tally calculating the neutrons flux between 0.5~keV and 10~keV. These two tallies create conflicting objectives for the optimization problem due to the moderation difference required to reach these energy ranges.  

\subsection{Ion-to-neutron Converter Problem}

\begin{figure}[!h]
    \centering
    \includegraphics[width=0.5\textwidth]{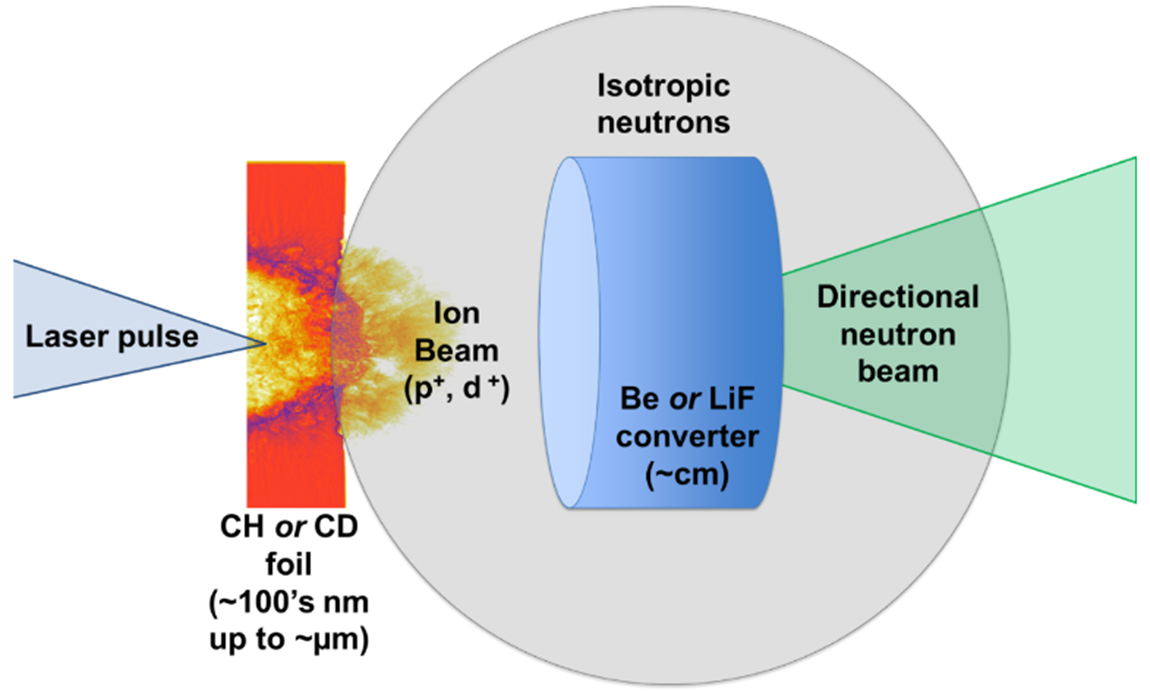}
    \caption{Illustration of the pitcher-catcher configuration used for laser-driven neutron production .}
    \label{fig:catcher_pitcher}
\end{figure}

The second design problem investigated in this paper is a ion-to-neutron converter design with high yield and high directionality. Directional neutron sources are widely used in various applications, including material analysis, imaging, isotope production, reactor optimization, and detecting fissile materials \cite{Anderson2001,Bowman1992,McDonald2007,VanDam1989}. The most effective short-pulse laser-driven neutron generation experiments use a two-stage “pitcher-catcher” configuration \cite{Huang2021, Horny2022}. The two-stage pitcher-catcher configuration is shown in Figure \ref{fig:catcher_pitcher}. In this configuration, a high-intensity laser (greater than $10^{18}$ W/cm²) is used to accelerate protons and deuterons from a thin deuteron enriched-plastic target. The deuteron enriched plastic target thickness typically ranges from 100 nanometers to 10 micrometers. This target is also referred to as the "pitcher". When the high-intensity laser pulse is incident on the pitcher, proton ($p^+$) and deuteron ($d^+$) ions located on this target are freed from the target and accelerated, creating an ion beam. This beam then interacts with a "catcher" target to produce neutrons. The catcher is usually composed of beryllium (Be) or lithium fluoride (LiF). The typical thickness of the catcher ranges from a millimeter to several centimeters. The (p,n) and (d,n) reactions in the catcher material leads to neutron production.

\begin{figure}[!h]
    \centering
    \includegraphics[width=0.7\textwidth]{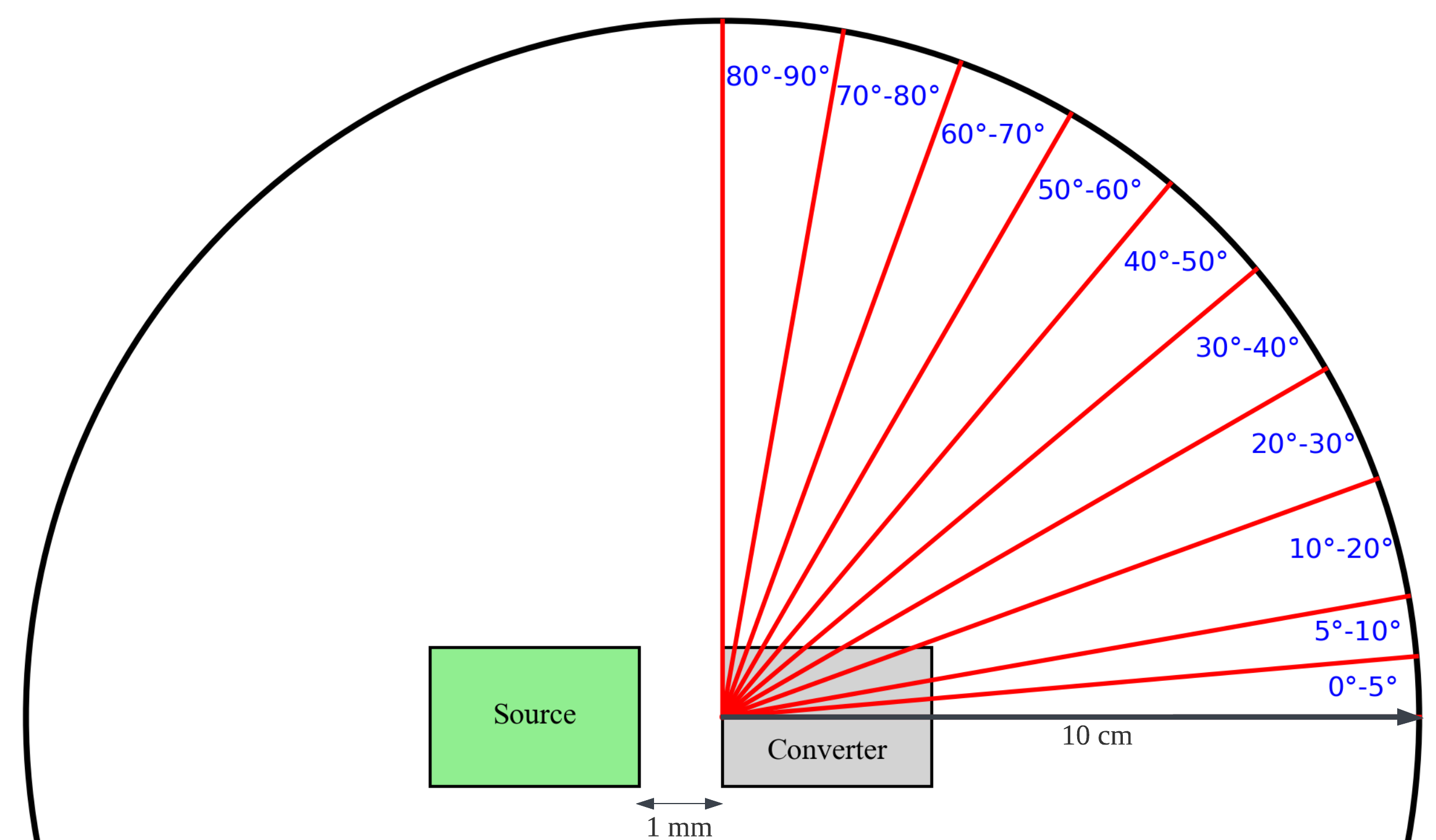}
    \caption{Simulated geometry for the ion-to-neutron converter problem. The ion-to-neutron converter dimensions are parametric in the problem, and are not drawn to size in this figure.}
    \label{fig:conv_prob_geom}
\end{figure}

In this study, the converter—also referred to as the catcher—is optimized to maximize both the average cosine of the emitted neutron distribution and the total neutron yield. The design is parameterized by two binary variables: the converter material and its geometric shape. The design is further parameterized by the base radius and height of both the cylindrical and conical structures. The positioning of the converter structure and problem geometry are given in Figure \ref{fig:conv_prob_geom}.

\begin{figure}[!h]
    \centering
    \includegraphics[width=0.3\textwidth]{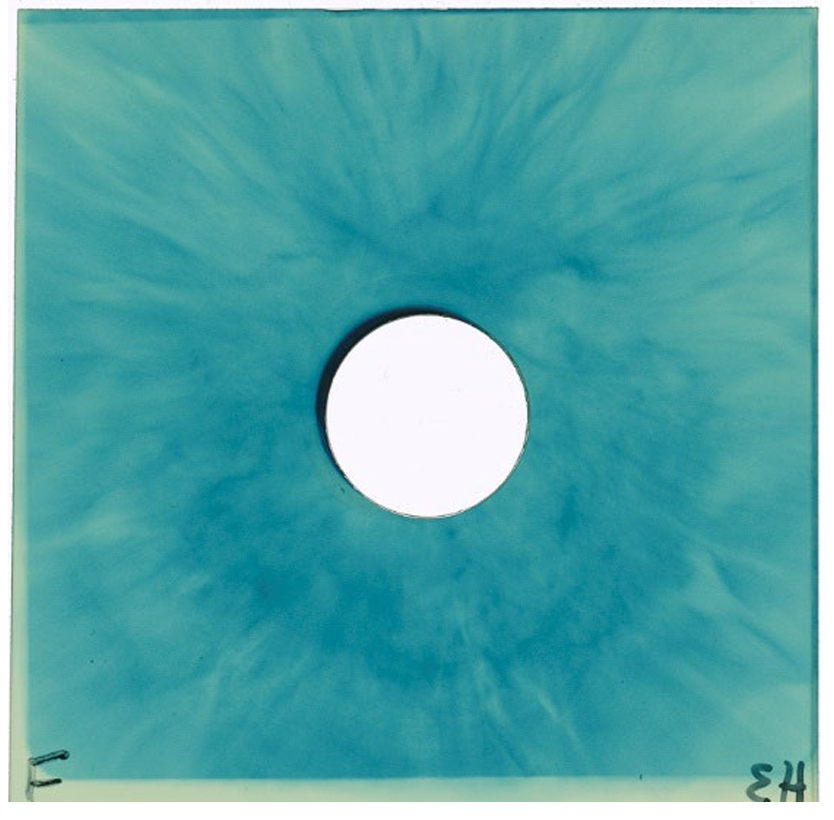}
    \caption{The radiochromatic film signal after the incident proton-deuteron pulse from the deuteron-enriched hydrocarbon film pitcher.}
    \label{fig:radiochromatic_film}
\end{figure}

To establish the ion beam properties for this study, we adopt experimental results obtained atthe OMEGA EP laser facility \cite{bohnen2008omega}. The proton-deuteron source was generated using a kilojoule-class beamline operated at 500 J with 0.7 ps compression, focused such that 80\% of laser intensity was contained within approximately 15 micron radius \cite{multiprobe}. This laser was incident upon a 700 nm thick CD (deuterated plastic) film. 
The proton and deuteron spatial distribution were been measured  using radiochromatic films located beyond the pitcher system. The signal distribution across the radiochromatic film in Figure \ref{fig:radiochromatic_film} was used for creating the directional distribution of the proton and deuteron sources within the MCNP simulations \cite{multiprobe}.  

The radiochromic film stack used for recording ion detections was positioned 10 cm from the pitcher structure. The film in Figure \ref{fig:radiochromatic_film}  had the highest resolution, and was selected for converting the ion detections into an angular profile of the ion beam. Films within the stack were spaced 0.8 cm apart, with the highest resolution film located 11.6 cm from the ion beam source. Given the film dimensions of 5 cm by 5 cm, the angle from the beam center to the film's farthest corner was calculated to be 0.11 radians. This value has been used for scaling the x and y axis values in the Gaussian function fitting in the next step. 

\begin{figure}[!h]
    \centering
    \includegraphics[width=0.4\textwidth]{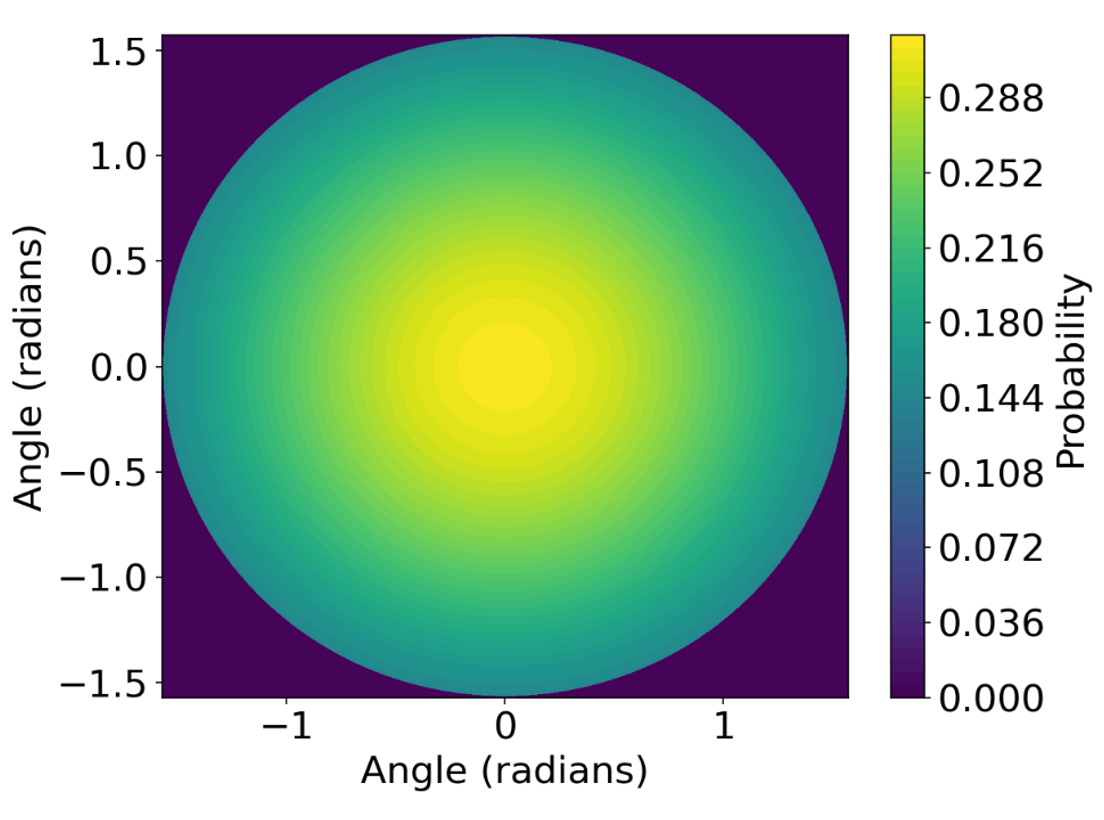}
    \caption{The angular ion source used in MCNP simulations. The steps in the figure represents the angular bins used in the MCNP simulations.}
    \label{fig:angular_dist}
\end{figure}

A two-dimensional Gaussian function was fit to the amplitude profile shown in Figure \ref{fig:angular_dist}. Prior to the analysis, the detection image was inverted to accurately capture the increased darkness corresponding to higher ion exposure. The fitted Gaussian function was normalized so that the integral of its magnitude over the range $[-\pi/2, \pi/2]$ equals one, ensuring an accurate representation of the angular distribution. The normalized Gaussian function is presented in Eq.\eqref{eq:prob1_src_gauss}, and the corresponding angular distribution of the ion source is illustrated in Figure \ref{fig:angular_dist}. 

\begin{equation}
    f(x,y) = 0.3164 \exp\!\left(-\frac{x^2+y^2}{2(1.2642)^2}\right)
    \label{eq:prob1_src_gauss}
\end{equation}

\begin{figure}[!h]
    \centering
    \includegraphics[width=0.4\textwidth]{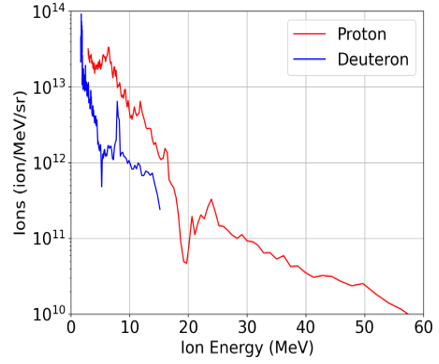}
    \caption{Energy distribution of protons and deuterons with a total of $1.1\times10^{13} p^+$ and $3.3\times10^{12} d^+$ ions.}
    \label{fig:energy_dist}
\end{figure}

In addition to the ion beam's angular distribution, its energy distribution is required for modeling. The ion energies were also derived from experimental data collected at the OMEGA EP. Figure \ref{fig:energy_dist} presents the ion spectra measured from a 700 nm thick CD foil irradiated with a 500 J, 0.7 ps pulse, as recorded using a Thomson Parabola spectrometer \cite{multiprobe} (Thompson parabola analysis described in \cite{Higginson2014}\cite{Huang2025}). The spectra were scaled to reflect the total number of protons and deuterons per shot—1.1×10$^{13}$ and 3.3×10$^{12}$, respectively. These yields indicate that the ion beam comprises approximately 77\% protons and 23\% deuterons. It is thought that the humidity on the surface of the plastic film is responsible for the the high proton yield in the ion beam. 

The energy distribution of the protons and deuterons in the ion beam is shown in Figure \ref{fig:energy_dist}. In our simulations, the MCNP source definitions specifying the proton and deuteron beam parameters were held constant across the catcher parameter space. This consistency ensures that any observed variations in the neutron energy distribution are attributable solely to differences in the catcher designs rather than fluctuations in the ion source characteristics.

Separate MCNP simulations were performed for protons and deuterons, using identical geometries and tally uncertainty cutoff values in each case. After the simulations, the resulting tallies were combined by weighting them according to the relative percentages of protons and deuterons present in the experimental ion beam measurement. The final, weighted tally values are reported on a per-laser-shot basis, ensuring that the simulation results accurately reflect the experimentally observed ion beam composition. Due to limitations in available data, a major assumption was made whereby the angular and energy dependencies of the ions were neglected in both the dataset creation and the MCNP simulations.

To analyze the angular distribution of neutrons behind the converter, a spherical tally surface of 10 cm radius was centered on the converter’s incident surface (i.e., where the ion beam first contacts the catcher). This spherical surface was then partitioned into ten distinct conical sectors, each corresponding to a specific polar angle range measured from the central axis normal to the converter surface. Specifically, the first sector covers 0° to 5°, the second 5° to 10°, the third 10° to 20°, and so on, up to 80° to 90°. By tallying neutrons crossing each conical sector, we capture the angular dependence of the neutron flux and energy distribution. This approach ensures that the influence of the converter geometry on neutron scattering is resolved across multiple angular intervals, facilitating a detailed comparison of directional neutron emission characteristics. 

Simulation tally values are used for establishing two optimization objectives of the problem. The first objective is derived by computing a weighted average cosine of the neutron angular distribution. The second objective is defined as the total neutron yield per laser pulse. This is the total number of neutrons observed on the spherical tally surface when the incident ion beam, as described in the previous section, interacts with the converter structure characterized by its selected height, base radius, shape, and material parameters. The calculation of the objective values from the tally results are explained in the next section.

\section{Optimization Methodology}
\label{sec:method}

The previous section defines the engineering problems that will be analyzed in this study using Monte Carlo simulations. These problem definitions are used for creating the dataset for surrogate model training. The input and output parameters of the MCNP simulations, the surrogate model data, surrogate model parameters, and optimization parameters applied on the surrogate models are explained in this section. 

\subsection{Dataset Input and Output Parameters}
\label{sec:met_inp_out}

\subsubsection{The Moderator Design Problem Parameters}
In the moderator design problem, the thicknesses of the moderator regions are the primary input parameters. Each material layer’s thickness was varied in the MCNP simulations to generate a comprehensive set of input files. The initial methodology was to present 3 input parameters for each layer's thickness in the moderator-reflector structure. However, the effect of the Pb layer inevitably saturates since the amount of the neutrons in the Pb layer decreases exponentially. To show this condition, the uncollided neutron flux in the one dimensional neutron transport problem is given in Eq.\eqref{eqn:1d_neut_atten} as an example. Using the saturated Pb layer thickness for all simulations removes an input parameter and decreases the computational cost of the data generation.

\begin{equation}
\label{eqn:1d_neut_atten}
 \phi(x, \mu) = \phi_0 \exp\left(-\frac{\Sigma_t x}{\mu}\right), 
\end{equation}
where $\phi(x, \mu)$ is the uncollided flux in moderator, $\phi_0$ is initial flux, and $\Sigma_t$ is the total cross section

Preliminary simulations revealed that the effect of the Pb layer saturates between 2 cm and 10 cm for various Be and PE thicknesses. Therefore, a uniform Pb layer thickness of 10 cm was adopted for all subsequent simulations. The Be and PE layer thickness values were varied systematically within an input parameter grid: the Be layer thickness ranged from 0.003 cm to 0.09 cm in 30 linearly spaced steps of 0.003 cm, while the PE layer thickness ranged from 0.75 cm to 2.5 cm in 25 linearly spaced steps of 0.03 cm. 

The objective functions for optimization are derived directly from the raw tally outputs of the MCNP simulation. These tallies measure the back-scattered neutron flux over a circular surface with a diameter of 0.8~cm, positioned 1~cm from the first moderator layer. The tallies record the neutron flux in units of cm$^2$/source particle, with one tally integrating over the 1–100~eV range and another over the 0.5–10~keV range. Thus, while the number of simulated particles (NPS) influences the statistical uncertainty, it does not alter the underlying measured values, which constitute the primary focus of this work.


Each of the simulations in the defined input parameter grid has been run with 5 different average "total tally" uncertainties. For example, if 1~\% minimal uncertainty is required, the simulation is stopped once the total flux uncertainty is smaller than that threshold. These tally uncertainty values have been selected as 10\%, 7.5\%, 5\%, 3\% and 1\%. The detectors in the energy range between 1 eV and 100 eV require more moderation in the moderator-reflector structure. The uncertainty value of this tally has been detected to be higher in the generated MCNP simulations. For this reason, the tally uncertainty cutoff has been set for the tally tracking the particles from 1 eV to 100 eV neutron energies. By using the described input parameter grid and the uncertainty values, a moderator design problem dataset has been created with 3750 samples, which will be used to train the surrogate model. The parameter range explored in this problem is summarized in Table \ref{tab:refl_params}.

\begin{table}[!ht]
\centering
\caption{Neutron Moderator Design Problem Input and Output Parameters.}
\begin{tabular}{|l|l|l|}
\hline
Parameter & Input/Output & Parameter Range \\ \Xhline{1pt}
Be Layer Thickness & Input & [0.003, 0.09] in 30 steps  \\ \hline
PE Layer Thickness & Input & [0.75, 2.5] in 25 steps \\ \hline
Tally Uncertainty & Input & 10\%, 7.5\%, 5\%, 3\%, 1\% \\ \hline
1-100 eV Neutron Flux Tally & Output & - \\ \hline
0.5-10 keV Neutron Flux Tally & Output & - \\ \hline
\end{tabular}
\label{tab:refl_params}
\end{table}

\subsubsection{The Ion-to-neutron Converter Design Problem Parameters}

In the ion-to-neutron converter problem, converter material, shape, height and radius have been used for creating the input space of the problem. Two materials, beryllium (Be) and lithium fluoride (LiF), were implemented in the MCNP simulations. Additionally, two distinct geometries were considered: a cylinder and a cone. For the cylindrical configuration, one of the circular base normals is oriented toward the ion beam while the opposite base is aligned with the directional neutron beam. In the conical configuration, the apex of the cone is directed at the ion beam, with the base aligned along the neutron beam direction. The maximum converter height for the simulations was set to the range of the highest-range protons in the problem—2.2 cm for 60 MeV protons in Be \cite{Karsch2010}. This value was thus adopted as the maximum height for both the cylindrical and conical converter geometries, while the maximum base radius was arbitrarily set at 1 cm. To systematically investigate the effects of these dimensions, an input parameter grid was established: the converter height was varied from 0.05 cm to 2.2 cm in 44 linearly spaced increments of 0.05 cm, and the base radius was varied from 0.1 cm to 1 cm in 10 linearly spaced increments of 0.1 cm. Each simulation has been created with for incident proton ions and incident deuteron ions. 

The output parameters (i.e., objectives) of the second problem are the average cosine of the emitted neutrons, and the total yield of the converter. To achieve the weighted average cosine objective, the neutron flux per laser pulse from each angular bin is multiplied by the cosine of the bin’s central angle. Since the conical surface area increases with higher cone apex angles, normalizing the flux—by dividing each tally by the total flux across all bins—effectively compensates for this geometric effect. The weighted cosine values are then summed to yield the average cosine. The value of the second objective, the total yield of the converter, is directly taken from the tally result of the simulations. This tally is placed around the catcher structure of the problem. The tally value signifies the number of detections on the spherical surface encompassing the pitcher and catcher structure.

\begin{figure}
    \centering
    \includegraphics[width=1\textwidth]{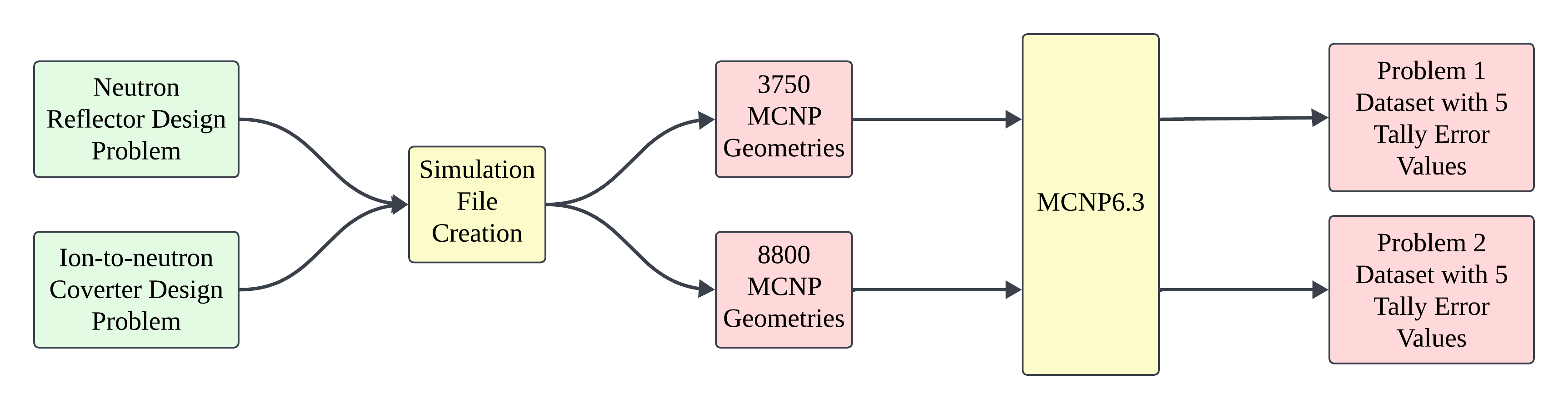}
    \caption{The flowchart summarizing the dataset creation.}
    \label{fig:data_flow}
\end{figure}

Each of the simulations in the defined input parameter grid was run using 5 different average tally uncertainty values: 5\%, 3.5\%, 2\%, 1\% and 0.2\%. The uncertainty value of the spherical neutron tally is the only tally surface in this problem. The tally uncertainty cutoff has been set for the spherical tally surface. By using the described input parameter grid and the uncertainty values, 17600 simulations with 8800 geometries (2 particle simulations per geometry) have been created for the surrogate model training. Two simulations with incoming p$^+$ and d$^+$ ions create a single converter behavior for an incoming ion beam. After running the MCNP simulations, the results are combined for 1.1×10$^{13}$ p$^+$ ions and 3.3×10$^{12}$ d$^+$ ions for each geometry. A converter design dataset has been created for 8800 geometries. The parameter range explored in the problem is summarized in Table \ref{tab:conv_params}. Figure \ref{fig:data_flow} illustrates the dataset creation steps.

\begin{table}[!ht]
\centering
\caption{Ion-to-neutron Converter Design Problem Input and Output Parameters}
\begin{tabular}{|l|l|l|}
\hline
Parameter & Input/Output & Parameter Range \\ \Xhline{1pt}
Converter Shape & Input & Cylinder, Cone \\ \hline
Converter Material & Input & Be, LiF \\ \hline
Shape Height & Input & [0.05, 2.2] in 44 steps \\ \hline
Shape Base Radius & Input & [0.1, 1] in 10 steps \\ \hline
Tally Uncertainty & Input & 5\%, 3.5\%, 2\%, 1\%, 0.2\% \\ \hline
Average Cosine of Neutrons & Output & - \\ \hline
Total Neutron Yield per Laser Pulse & Output & - \\ \hline
\end{tabular}
\label{tab:conv_params}
\end{table}

\subsection{FNN Surrogate Model Training}
\label{sec:met_surrogates}

In the previous section, we prepared data comprising two distinct problems, each characterized by five levels of tally uncertainty. For each problem, the dataset was partitioned into five subsets based on these uncertainty levels, and a surrogate model was developed for each subset.

A grid search was employed to tune the hyperparameters of the surrogate models. This optimization approach was applied to the five models corresponding to the distinct tally uncertainty levels for both the neutron moderator design problem and the ion-to-neutron converter design problem. Table \ref{tab:conv_surr_params} summarizes the hyperparameter grid explored during this tuning process.

\begin{table}[ht]
\centering
\caption{Hyperparameter grid used in the tuning of the surrogate models.}
\begin{tabular}{|l|l|}
\hline
Model Parameters & Values \\ \hline
Number of Layers & 1, 4, 7, 10 \\ \hline
Neurons per Layers & 100, 400, 700, 1000 \\ \hline
Learning Rate & 1e-3, 4e-4, 1e-4 \\ \hline
Batch Size & 1, 2, 4 \\ \hline
\end{tabular}
\label{tab:conv_surr_params}
\end{table}

The hyperparameter tuning is performed for Keras/Tensorflow-based FNN surrogate models using fully-connected dense layers. Each model is built with hidden layers that employ the rectified linear unit (ReLU) activation function to capture non-linear relationships. Prior to training, both input features and target values are standardized using standard scaling, where the inputs have a mean of 0 and a standard deviation of 1. This improves convergence and overall model performance.

During hyperparameter tuning, the model is configured to train for a maximum of 200 epochs while employing an early stopping strategy. Early stopping monitors the test set Mean Squared Error (MSE) and terminates training if no improvement is observed over 20 consecutive epochs. The early stopping decreases the computational cost of the hyperparameter tuning process and prevents overfitting by halting training when no significant improvement is detected. As a result of using early stopping in the training process, the training of the models typically ended in fewer than 150 epochs. The performance of each model is measured using the $R^2$ (coefficient of determination) metric on a held-out test set. To manage the computational load of this exhaustive search, the process is parallelized, allowing simultaneous evaluation of many hyperparameter combinations.

Hyperparameter tuning is carried out independently for each of the five tally uncertainty data subsets corresponding to two different problems—yielding a total of ten distinct hyperparameter sets. For each subset, a grid search is performed over various combinations of the number of layers, neurons per layer, batch sizes, and learning rates, with training executed in parallel to enhance efficiency. The optimal hyperparameter set for each dataset is then used to export the best performing FNN surrogate models, which are subsequently employed in the NSGA-III optimization process as detailed in the following subsection.

\subsection{NSGA-III Optimization of Surrogate Models}
\label{sec:met_optimization}


After the surrogate models have been trained and the best hyperparameters identified, the next step is to employ these models within an NSGA-III optimization framework. This process is executed separately for each of the five surrogate models across two distinct problems, resulting in a total of ten independent optimization runs. The following subsection outlines the methodology and key parameters involved in the NSGA-III optimization process.

The pre-trained Keras models serve as surrogate evaluators. Instead of directly performing costly high-fidelity MCNP simulations, these models predict objective function values for a given set of input parameters, significantly reducing computational overhead. The two optimization problems described in Section \ref{sec:problems} are considered for representing with FNN surrogate models. Each problem is defined by its own set of design variables, objective functions, and constraints. The surrogate models are tailored to capture the specific characteristics of the data corresponding to each problem.

NSGA-III \cite{deb2014nsga3} is a state-of-the-art evolutionary algorithm designed to solve many-objective optimization problems. It employs reference directions to guide the selection process, ensuring a diverse spread of solutions along the Pareto-front. The optimization begins with the generation of a diverse initial population. This is achieved by sampling within the predefined bounds of the design variables, ensuring the proper exploration of the solution space. The number of candidate solutions maintained in each generation is critical. It is chosen based on the problem complexity and the number of objectives to ensure sufficient diversity. The maximum number of evolutionary cycles is defined to control the computational budget. The algorithm iterates through these generations until convergence criteria are met. Genetic operators—crossover and mutation—are applied to the population to generate new candidate solutions. Their probabilities are set to balance exploration (searching new areas) and exploitation (refining existing solutions). A set of evenly distributed reference directions is established. These directions guide the selection process by associating each candidate with a specific reference point, thereby maintaining a well-distributed Pareto-front. The optimization process concludes when either the maximum number of generations is reached or the solution improvements fall below a specified threshold, indicating convergence.

In each generation, the algorithm applies selection, crossover, and mutation to create offspring, combining traits of the current population to explore new regions of the design space. Each candidate solution is evaluated using the corresponding surrogate model. This evaluation predicts the performance metrics for each candidate, serving as a basis for multi-objective ranking. Candidates are ranked based on Pareto dominance, and the reference direction mechanism ensures that the selected solutions represent a diverse set of trade-offs between objectives. Upon termination, the algorithm outputs a Pareto-optimal set of solutions. These represent the best trade-offs found between competing objectives, offering valuable insights for decision-making.

By applying the NSGA-III optimization separately for each of the five data subsets (corresponding to different tally uncertainty levels) across two problems, the methodology addresses a variety of scenarios. This integrated methodology of surrogate modeling and NSGA-III optimization enables efficient exploration of the design space in both problems.

\begin{figure}
    \centering
    \includegraphics[width=1\textwidth]{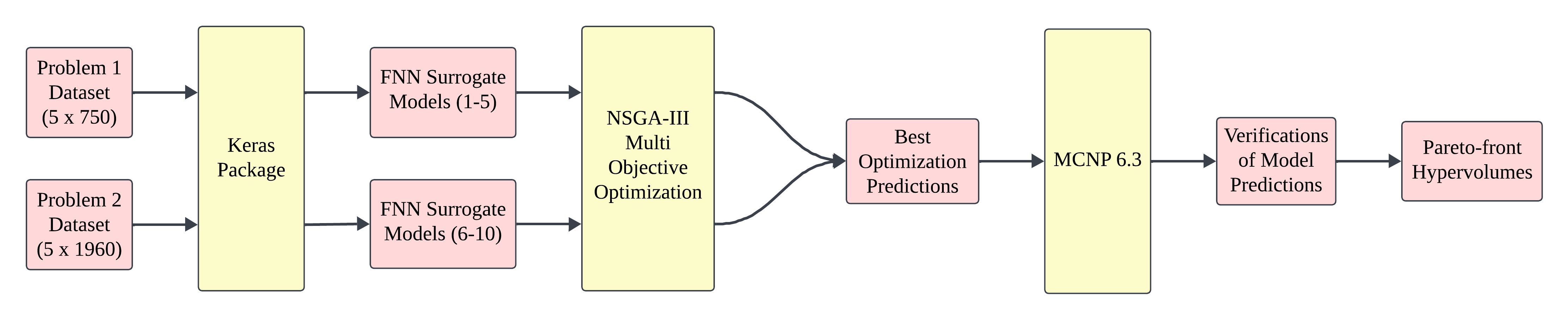}
    \caption{The flowchart summarizing the calculation of hypervolume loss for each FNN surrogate model created with different MCNP tally uncertainty values in the training data.}
    \label{fig:data_to_hypervol}
\end{figure}

Once the NSGA-III optimization has produced its results based on the hyperparameter-tuned surrogate models, the best-performing individuals—those with the optimal input parameters—are extracted. These parameters are then used as inputs for high-fidelity verification simulations using MCNP code. This step evaluates the validity of the surrogate-based predictions by checking them against rigorous, computationally expensive simulation results, ensuring that the surrogate models’ outcomes are verified. The research methodology to create the surrogate models, optimization predictions and hypervolume results are given in Figure \ref{fig:data_to_hypervol}.

In the verification phase, the results from the MCNP simulations are compared against the simulations conducted using the lowest uncertainty dataset (i.e., the 1\% tally uncertainty dataset for the moderator design problem and 0.2\% tally uncertainty dataset for the converter problem), which is treated as the ground truth. The loss of Pareto-front is reported with increased tally uncertainty limit. The comparison is based on the normalized hypervolume metric—a measure that quantifies the volume of the objective space dominated by the Pareto-front. For both optimization problems, the worst-case point (nadir point) is defined as the combination of the worst observed objective values in the dataset. Pareto-front hypervolumes are computed with respect to this reference point. 2D hypervolume calculation equation is given in Eq.\eqref{eq:2hypervol}. By evaluating what percentage of the ground truth hypervolume is achieved by the best individuals from the NSGA-III run, the methodology provides a clear and quantitative measure of how closely the surrogate-driven optimization approximates the performance of the most accurate simulations. This analysis not only validates the efficacy of the surrogate models but also offers insights into potential areas for further refinement.

\begin{equation}
\text{HV-2D} = \frac{1}{2} \left| \sum_{k=1}^{N-1} \left( 
y_1^{(R)}(y_2^{(k)} - y_2^{(k+1)}) + 
y_1^{(k)}(y_2^{(k+1)} - y_2^{(R)}) + 
y_1^{(k+1)}(y_2^{(R)} - y_2^{(k)}) 
\right) \right|
\label{eq:2hypervol}
\end{equation}
where $(y_1^{(R)},y_2^{(R)})$ is the reference point, the points $(y_1^{(1)},y_2^{(1)})$ to $(y_1^{(N)},y_2^{(N)})$ are the sorted 2D Pareto-front, and $N$ is the number of Pareto-front individuals.

\subsection{Computing Resources}
\label{sec:met_comp_res}

To maximize computational efficiency during dataset generation, this study used five distinct clusters at Los Alamos National Laboratories to maximize computational efficiency during the dataset generation stage.  For the neutron moderator design problem, the MCNP simulation input files were distributed across four clusters—Bobcat, Thundercloud, Cougar, and Ocelot (in total $\sim$3500 cores)—located at LANCE and the Lujan Center within the P-2 group, which operate under the TurbOS environment~\cite{hoonify2025turbos}. Meanwhile, the ion-to-neutron converter design problem was processed on the Midnight cluster of the Nuclear Engineering and Nonproliferation (NEN) Division. This distributed approach ensures robust and scalable performance by leveraging the specialized resources available at each cluster.

The surrogate model hyperparameter tuning in this research used an internal server at the University of Michigan. The grid search was performed using a multithreading calculation of 144 threads (72 cores), and successfully completed over a span of 6 hours. For the FNN surrogate model training, we employed TensorFlow 2.18.0 and Keras 3.9.0 to train the models. The optimization of the FNN surrogate models have been done by using NEORL 1.8 \cite{radaideh2023neorl} Python package. We conducted the hyperparameter tuning and surrogate model training on the same internal server, which is equipped with two AMD EPYC 9654 processors, each providing 96 cores operating at 2.4–3.7 GHz, resulting in a total of 192 cores and 384 threads. Additionally, the server features four NVIDIA RTX 6000 Ada Generation GPUs and 1536 GB of DDR5 RAM. 

\section{Results and Discussions}
\label{sec:results}

This section is organized into four parts. First, we present the initial MCNP simulation results used for data generation. Second, we discuss the training outcomes of the FNN surrogate model. Third, we detail the optimization performed on the surrogate models. Finally, we verify the optimization results with additional MCNP simulations.

\subsection{Dataset Generation}

\begin{figure}[!h]
    \centering
    \includegraphics[width=0.55\textwidth]{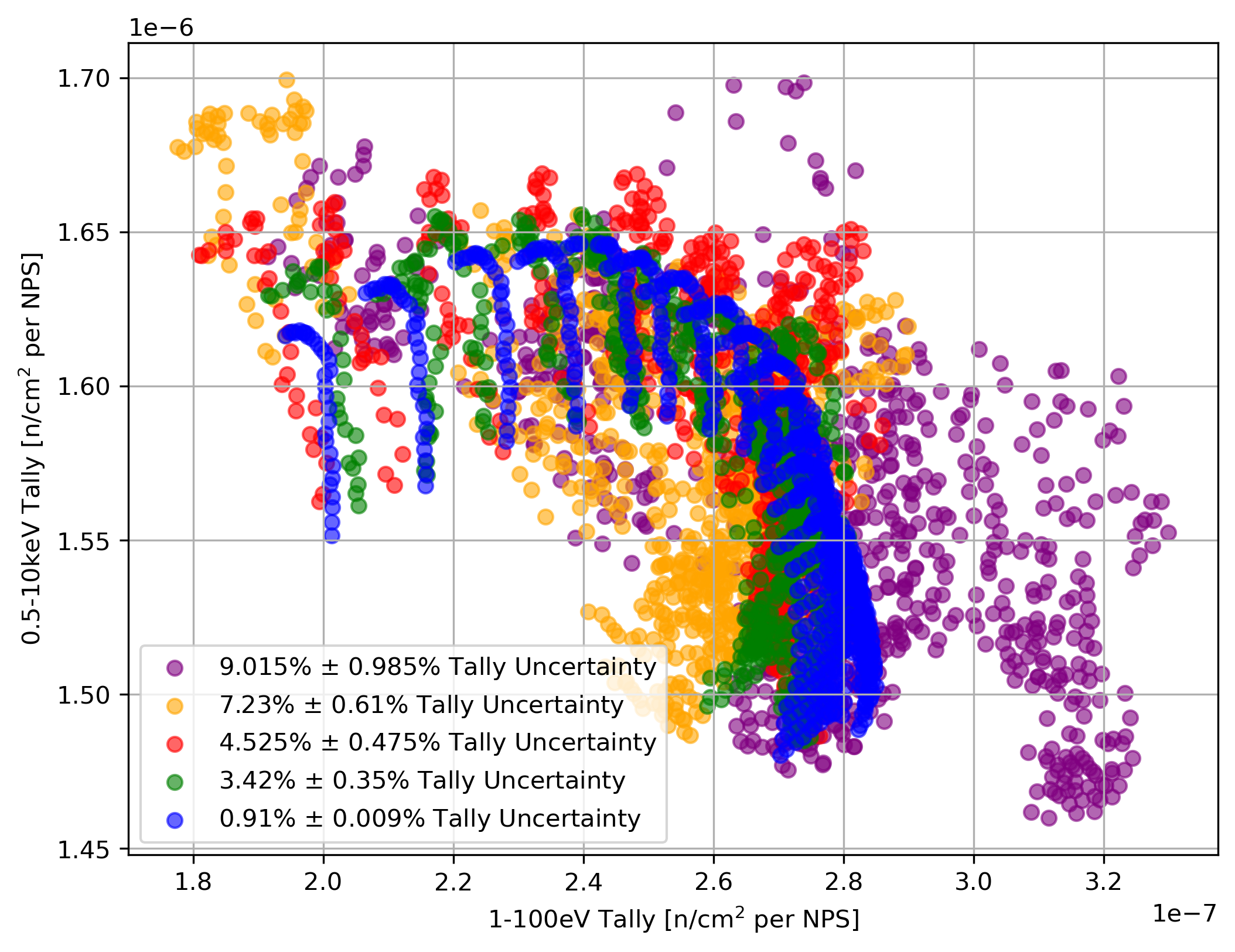}
    \caption{The data obtained from the MCNP simulation results of the neutron moderator design problem.}
    \label{fig:refl_data_plot}
\end{figure}

\begin{figure}[!h]
    \centering
    \includegraphics[width=0.55\textwidth]{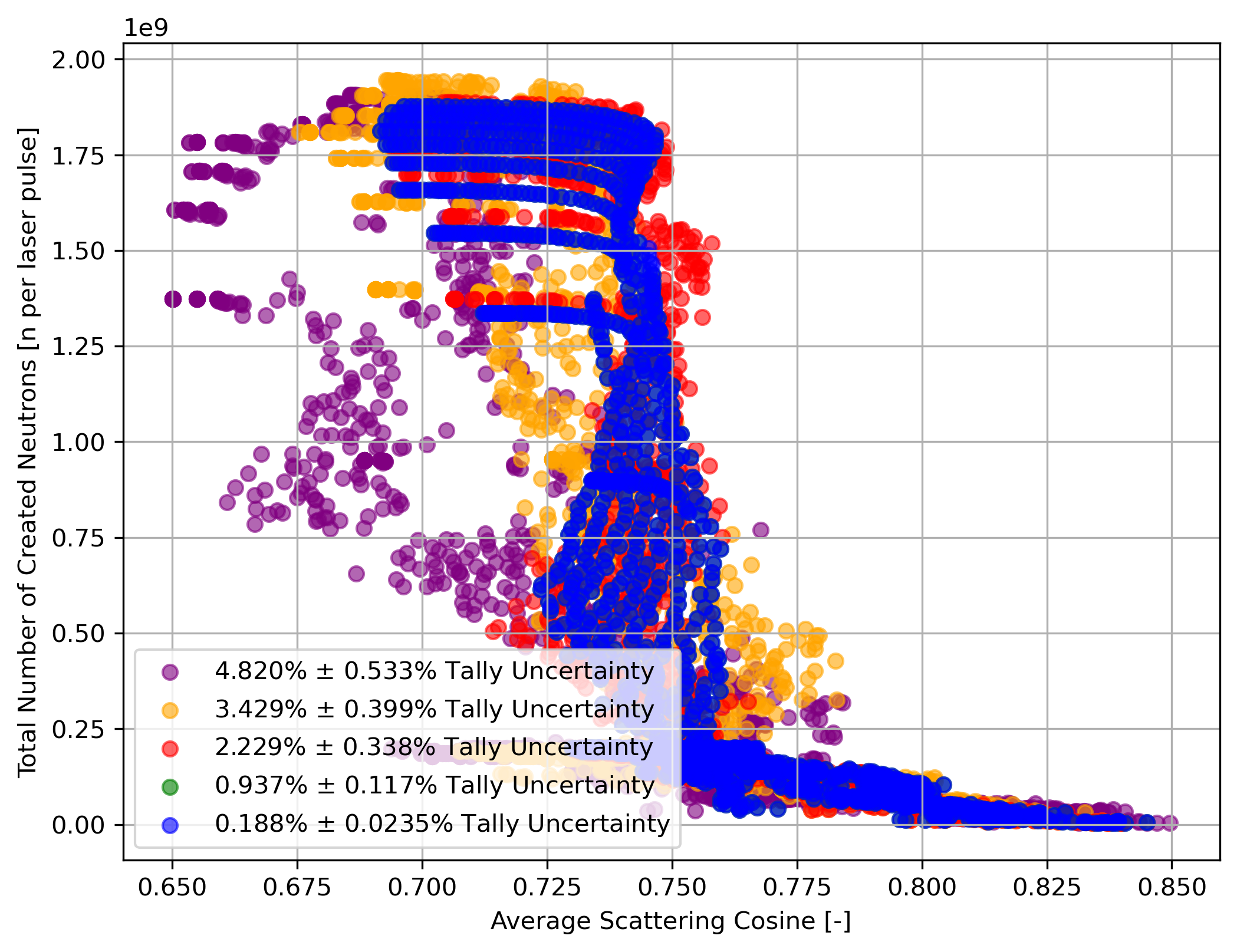}
    \caption{The data obtained from MCNP simulation results of the ion-to-neutron converter design problem.}
    \label{fig:conv_data_plot}
\end{figure}

The dataset described in Section \ref{sec:met_inp_out} comprises two distinct problems across five different tally uncertainty levels. Specifically, the neutron moderator design problem yields a total number of 3,750 data points, while the ion-to-neutron converter problem contributes a total number of 8,800 data points. To construct these datasets, MCNP simulations were executed using the corresponding input parameters for each data point, and the resulting tally values were extracted. Figure \ref{fig:refl_data_plot} and Figure \ref{fig:conv_data_plot} provide visual illustrations of how the tally values for the neutron moderator design problem and ion-to-neutron converter design problem evolve under different levels of tally uncertainty. 

Each colored cluster in the plot corresponds to a specific tally uncertainty level, revealing both the spread and overall displacement of data points in the two-dimensional tally space. Notably, when the tally uncertainty is small (around 1\%), the data points follow a relatively well-defined curve. This curve can be interpreted as a Pareto-front, representing the trade-off between the 1–100 eV tally and the 0.5–10 keV tally of Figure \ref{fig:refl_data_plot} and the trade-off between the total number of neutrons created and average neutron scattering cosine in Figure \ref{fig:conv_data_plot}. In contrast, as the error level increases (beyond 5\%), the clusters become more diffuse and the Pareto-front effectively disappears, obscured by the larger scatter in the data. Please note that the difference between 1\% and 0.2\% data is very small in the ion-to-neutron converter design problem data. The difference can only be detected in smaller scales. 

One way to interpret this increasing scatter is to consider how stochastic fluctuations in neutron transport simulations can inflate or distort the observed tallies. At lower error levels, the effects of random sampling in MCNP simulations remain small enough that the underlying physics-driven trends are still clearly visible. The tallies align more closely with deterministic expectations, making it possible to delineate trade-offs between different energy ranges. However, at higher error levels, the random fluctuations are significant enough to mask these trends, making the dataset appear “noisier.” This noise can lead to tallies that deviate substantially from their nominal converged mean values, thus broadening the distribution and merging regions of the design space that would otherwise be distinct.

Additionally, Figure \ref{fig:refl_data_plot} and Figure \ref{fig:conv_data_plot} highlight how tally uncertainties can cause both a systematic shift in the data and an expansion of its spread. The systematic shift might appear as a slight shift in the centroid of each cluster—meaning that on average, the tallies at a higher error level can be offset from those at a lower error level. Meanwhile, the increased spread indicates that even points with similar input parameters can exhibit noticeably different tally results, making it difficult to identify clear performance optima or trade-off boundaries. For tasks such as optimization or surrogate model training, this ambiguity complicates the search for best-performing designs as we will be noticing later on. When building surrogate models, the broader scatter requires more robust fitting techniques to avoid overfitting spurious patterns or underfitting legitimate trends.

From a multi-objective optimization standpoint, the disappearance of the Pareto-front at higher error levels is particularly important. A clear Pareto-front, as seen in the 1\% error dataset in Figure \ref{fig:refl_data_plot}, allows designers and analysts to select trade-off solutions based on their specific performance criteria in different energy ranges. However, once the front becomes obscured, the optimization process may converge to suboptimal points or yield highly uncertain estimates of the objective functions. This highlights the practical implications of tally uncertainties: even if one is theoretically interested in optimizing multiple objectives (e.g., flux in different energy ranges), the accuracy and precision of the tallies can become the limiting factor in how well those objectives can be discerned and balanced.

To summarize the dataset generation with differing error percentages, Figure \ref{fig:refl_data_plot} and Figure \ref{fig:conv_data_plot} show the critical role of tally uncertainty in shaping the distribution of tally results. Low-error data reveals clear trade-offs and patterns that can guide design decisions but are expensive to generate, whereas high-error data that are not expensive obscure these trends, complicating the modeling and optimization tasks. The implications of this tally uncertainty for surrogate model construction are explored further in the following sections.

\subsection{Surrogate Model Creation}
\label{sec:surr_create}

The performances of the FNN surrogate models were evaluated using two regression metrics: Mean Squared Error (MSE) and the coefficient of determination ($R^2$). MSE quantifies the average squared difference between predicted and actual values, penalizing larger errors more severely, while $R^2$ measures the proportion of variance in the target variable explained by the model—with values closer to 1 indicating a better fit. These metrics are mathematically defined as follows:

\begin{align}
\text{MSE} &= \frac{1}{n} \sum_{i=1}^{n} (y_i - \hat{y}_i)^2, & \quad
R^2 &= 1 - \frac{\sum_{i=1}^{n} (y_i - \hat{y}_i)^2}{\sum_{i=1}^{n} (y_i - \bar{y})^2},
\label{eq:metrics}
\end{align}
where $n$ is the number of samples in the test set, $\hat{y}_i$ is the predicted value of sample $i$ by the surrogate model, $y_i$ is the true value for sample $i$, and $\bar{y}$ is the mean of the true values ($y_i$) for all samples from 1 to $n$. In training, each surrogate model was optimized by minimizing the test data MSE. For each model, 15\% of the data subset was reserved as the test set, and the model achieving the lowest test MSE during training was selected. After training, the models were further compared based on their test set $R^2$ scores. Hyperparameter tuning was performed over the input parameter grid described in Table \ref{tab:conv_surr_params}. The parameter sets that produced the highest test set $R^2$ scores were deemed optimal. For the neutron moderator design problem, the best model parameters are reported in Table \ref{tab:refl_surr_results}, while those for the ion-to-neutron converter design problem are listed in Table \ref{tab:conv_surr_results}. 

\begin{table}[ht]
\centering
\caption{Tuned surrogate model parameters for neutron moderator design problem.}
\begin{tabular}{|l|l|l|l|l|l|}
\hline
Tally Uncertainty & Number of Layers & Neurons per Layers & Learning Rate & Batch Size & Model R$^2$ Score \\ \hline
1\% & 4 & 400 & 1e-4 & 2 & 0.999724 \\ \hline
3\% & 4 & 1000 & 1e-4 & 4 & 0.994461 \\ \hline
5\% & 4 & 700 & 1e-4 & 4 & 0.997316 \\ \hline
7.5\% & 4 & 1000 & 1e-4 & 2 & 0.995442 \\ \hline
10\% & 4 & 400 & 4e-4 & 4 & 0.992057 \\ \hline
\end{tabular}
\label{tab:refl_surr_results}
\end{table}

\begin{table}[ht]
\centering
\caption{Tuned surrogate model parameters for ion-to-neutron converter design problem.}
\begin{tabular}{|l|l|l|l|l|l|}
\hline
Tally Uncertainty & Number of Layers & Neurons per Layers & Learning Rate & Batch Size & Model R$^2$ Score \\ \hline
0.2\% & 7 & 700 & 1e-4 & 4 & 0.997795 \\ \hline
1\% & 4 & 700 & 1e-4 & 2 & 0.997810 \\ \hline
2\% & 4 & 1000 & 1e-4 & 2 & 0.994906 \\ \hline
3.5\% & 4 & 400 & 1e-4 & 2 & 0.992572 \\ \hline
5\% & 7 & 700 & 1e-4 & 4 & 0.991109 \\ \hline
\end{tabular}
\label{tab:conv_surr_results}
\end{table}

Table \ref{tab:refl_surr_results} and Table \ref{tab:conv_surr_results} summarize the tuned hyperparameters for the FNN surrogate models developed for the selected design problems. Each row corresponds to a different level of tally uncertainty in the training data, highlighting how the inherent noise influences the model architecture and training settings. 

For the selected design problems, most surrogate models were configured with 4 and 7 layers, suggesting that this depth was sufficient to capture the underlying simulation behavior across different noise levels. However, the number of neurons per layer varied—ranging from 400 to 1000—which indicates that some models required a higher capacity to effectively learn from more complex or noisier data. The learning rate was maintained at 1e-4 for most cases, with a slight increase to 4e-4 for a single case of 10\% uncertainty model. Batch size adjustments (from 2 to 4) further reflect the fine-tuning process aimed at optimizing the convergence behavior of each model based on the quality of the input data. The high R$^2$ scores (ranging from 0.991 to 0.999) demonstrate that the surrogate models generalize extremely well to the test data, faithfully replicating the behavior of the original MCNP simulations. This near-perfect performance confirms that the models are capable of capturing the complex input-output relationships despite the varying levels of inherent tally uncertainty, and the models are perfectly generalized for the problems and ready to provide accurate predictions.

\begin{figure}[!htb]
    \centering
    \begin{subfigure}{0.5\textwidth}
        \centering
        \includegraphics[width=\linewidth]{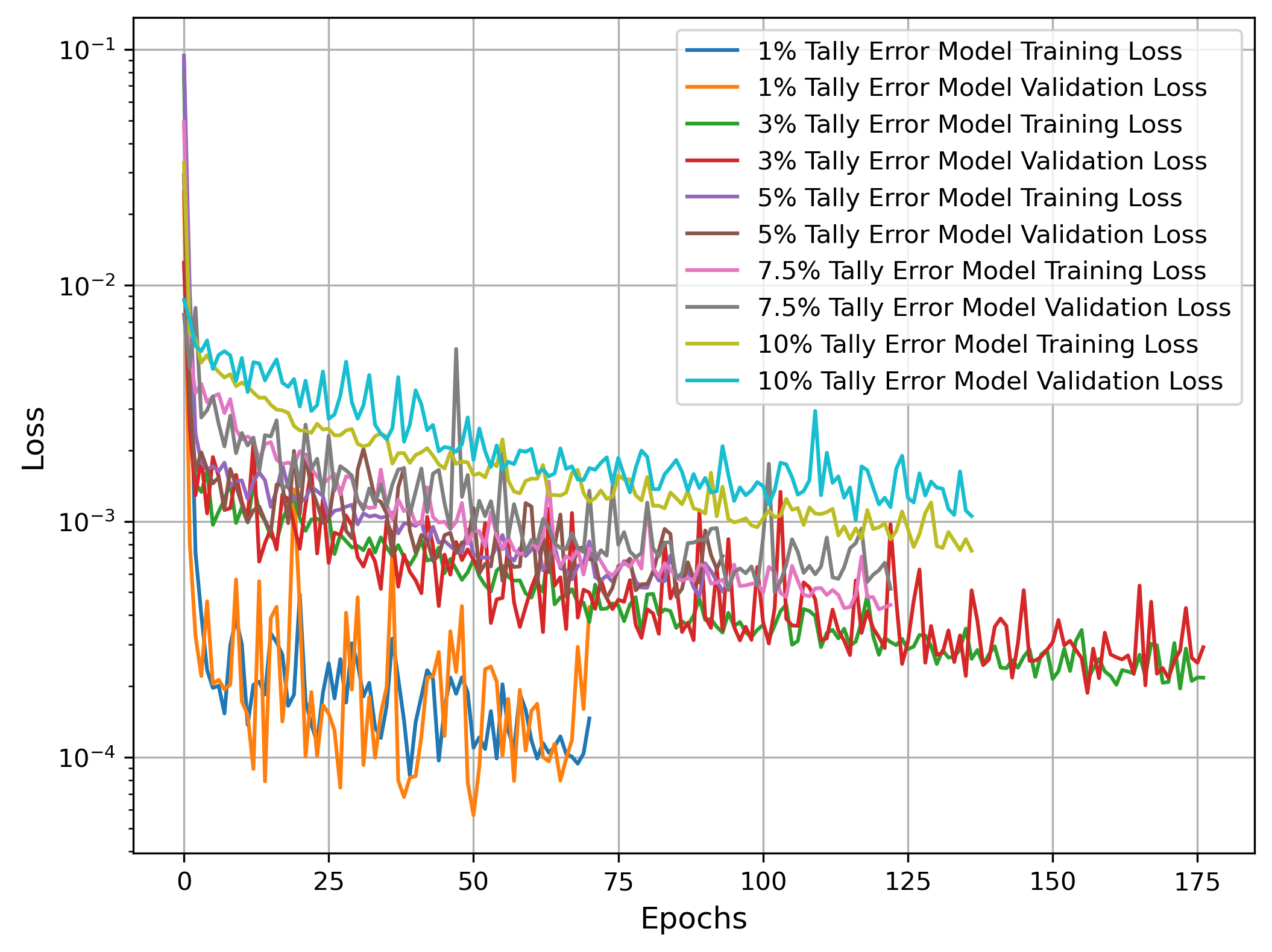}
    \end{subfigure}%
    \hfill
    \begin{subfigure}{0.5\textwidth}
        \centering
        \includegraphics[width=\linewidth]{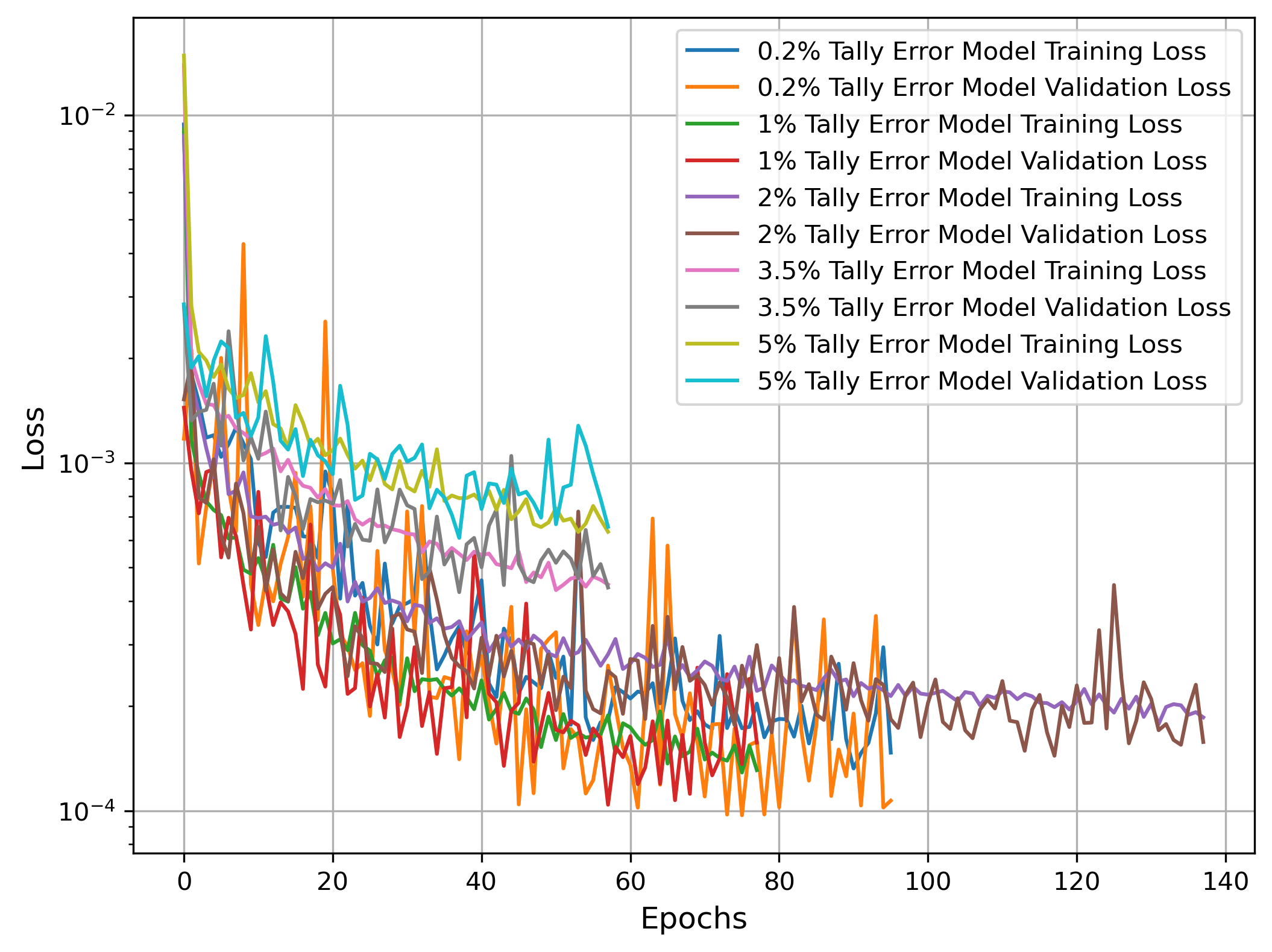}
    \end{subfigure}
    \caption{The validation MSE metrics of FNN surrogate models trained for neutron moderator design problem (left) and ion-to-neutron converter design problem (right).}
    \label{fig:surr_val_plot}
\end{figure}

\begin{figure}[!h]
    \centering
    \includegraphics[width=1\textwidth]{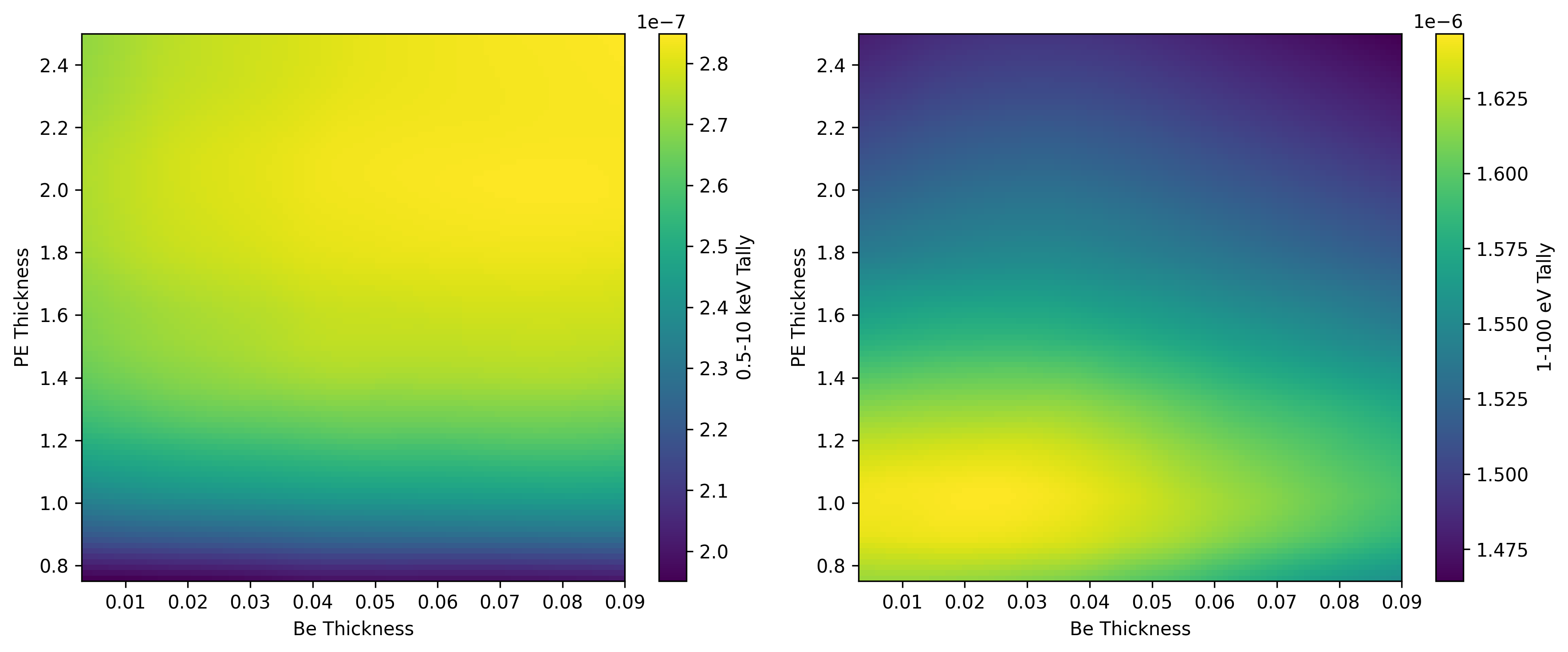}
    \caption{The input-output relation of a selected FNN surrogate model (trained with neutron moderator design problem data with 1\% tally uncertainty).}
    \label{fig:refl_surr_input_output}
\end{figure}

While the surrogate models accurately predict the MCNP simulation outputs, they inherently incorporate the tally uncertainty present in their training and test data. The observed variations in hyperparameters across different uncertainty levels underscore the importance of careful hyperparameter tuning. After identifying the optimal hyperparameters, the models were retrained to allow for a more in-depth analysis of their characteristics. Figure \ref{fig:surr_val_plot} presents the evolution of training and validation MSE metrics for each surrogate model, clearly showing that the minimum MSE achieved is inversely related to the error percentage of the training data. This trend is further supported by the R$^2$ metrics presented in Tables \ref{tab:refl_surr_results} and \ref{tab:conv_surr_results}. Notably, no correlation was found between the number of training epochs and the error percentage of the training data.

The surrogate model developed for the neutron moderator design problem using a 1\% tally uncertainty was chosen as a representative example; its input-output relationship is depicted in Figure \ref{fig:refl_surr_input_output}. In the neutron moderator design problem, all surrogate models show a similar input-output relationship in the problem with variations dependent on their inherent training data tally uncertainty percentage. The 10 refined FNN surrogate models, as summarized in Tables \ref{tab:refl_surr_results} and \ref{tab:conv_surr_results}, serve as the basis for the subsequent surrogate model optimization phase.

\subsection{Surrogate-driven Multi-objective Optimization}
\label{sec:surr_opt}

\begin{figure}[!h]
    \centering
    \includegraphics[width=1\textwidth]{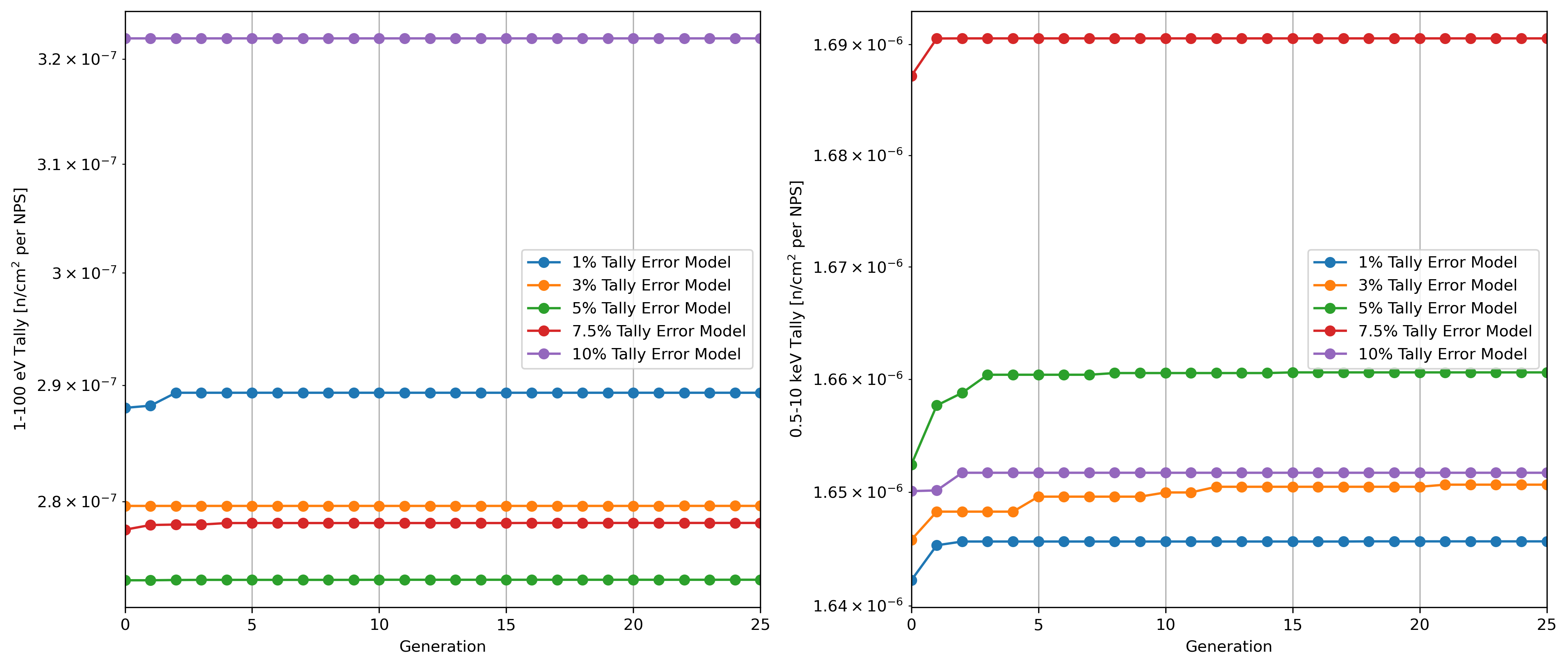}
    \caption{Change of the \textbf{best objective value} discovered during optimization of the neutron moderator design problem using its trained surrogate models. Only the first 25 generations are plotted since the next 75 generations do not show improvement in the best objective values. The discovered highest value of the 1-100 eV tally value does not change between iterations since the objective space is simple (as given in Figure \ref{fig:refl_surr_input_output}) and it converges very quickly.}
    \label{fig:refl_opt_fitness_change}
\end{figure}

\begin{figure}[!h]
    \centering
    \includegraphics[width=1\textwidth]{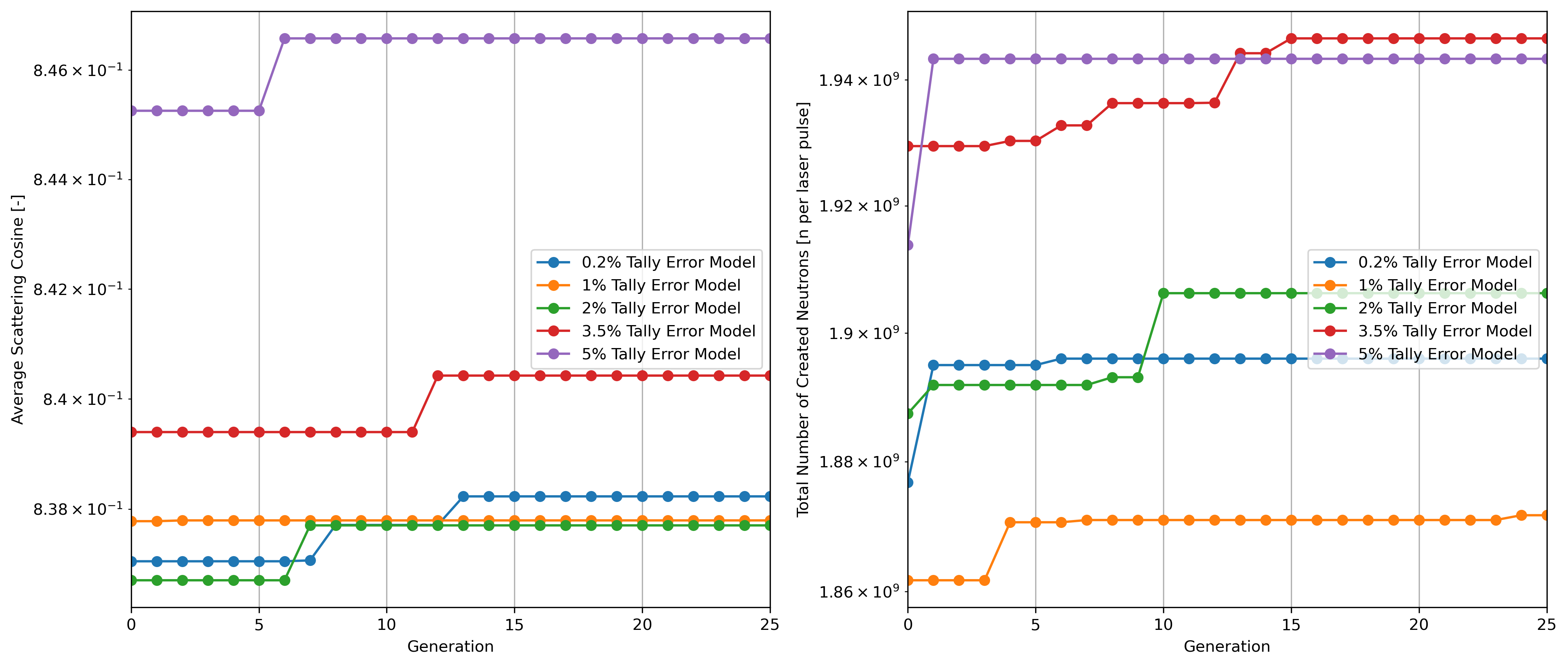}
    \caption{Change of the \textbf{best objective value} discovered during optimization of the ion-to-neutron design problem using its trained surrogate models. Only the first 25 generations are plotted since the next 75 generations do not show improvement in the best objective values.}
    \label{fig:conv_opt_fitness_change}
\end{figure}

To further explore the design space and identify optimal trade-offs between conflicting objectives, we employed the NSGA-III evolutionary algorithm using the NeuroEvolution Optimization with Reinforcement Learning (NEORL) \cite{radaideh2023neorl} package. The optimization process was configured with the following parameters: a population size of 100, a two-point crossover mode with an alpha value of 0.5 and a crossover probability of 0.7, a mutation probability of 0.2, and mutation limits set between 0.01 and 0.5. The algorithm was executed for 100 generations. Figures \ref{fig:refl_opt_fitness_change} and \ref{fig:conv_opt_fitness_change} illustrate the performance of the NSGA-III multiobjective optimization routine using two semi-logarithmic subplots. Figure \ref{fig:refl_opt_fitness_change} shows the change of fitness values in each generation for the neutron moderator design problem. Figure \ref{fig:conv_opt_fitness_change} shows the change of fitness values in each generation for the ion-to-neutron converter problem. In each of these figures, all FNN surrogate models are the models trained with different tally uncertainty values in the same problem. The left subplot displays the evolution of the best uncovered objective 1 values, while the right subplot shows the corresponding progression for objective 2 values. Each subplot includes five curves that represent the optimization runs performed on different surrogate models, with the \textbf{best objective values found in every generation} are plotted. With a population size of 100, all routines converge after 25 generations. The variation in the converged values reflects the differences in the fitness functions of the surrogate models. Notably, all routines achieve maximization of their respective objective values within these generations, demonstrating the efficiency of the NSGA-III algorithm on these problems. 

The five surrogate models in Figure \ref{fig:refl_opt_fitness_change} are applied to the neutron moderator design problem, and five surrogate models in Figure \ref{fig:conv_opt_fitness_change} are applied to the ion-to-neutron converter design problem. Yet each of the models are trained on datasets with distinct tally uncertainty rates. The maximum achievable objective value of each model is influenced by both the inherent best performance and the associated tally uncertainty at that optimum point. Consequently, the optimal objective values differ among the FNN surrogate models. Even though the surrogate models address the same problem, the best solutions they produce vary as shown in Figure \ref{fig:refl_opt_fitness_change} and \ref{fig:conv_opt_fitness_change}.

\begin{figure}[!h]
    \centering
    \begin{subfigure}{0.5\textwidth}
        \centering
        \includegraphics[width=\linewidth]{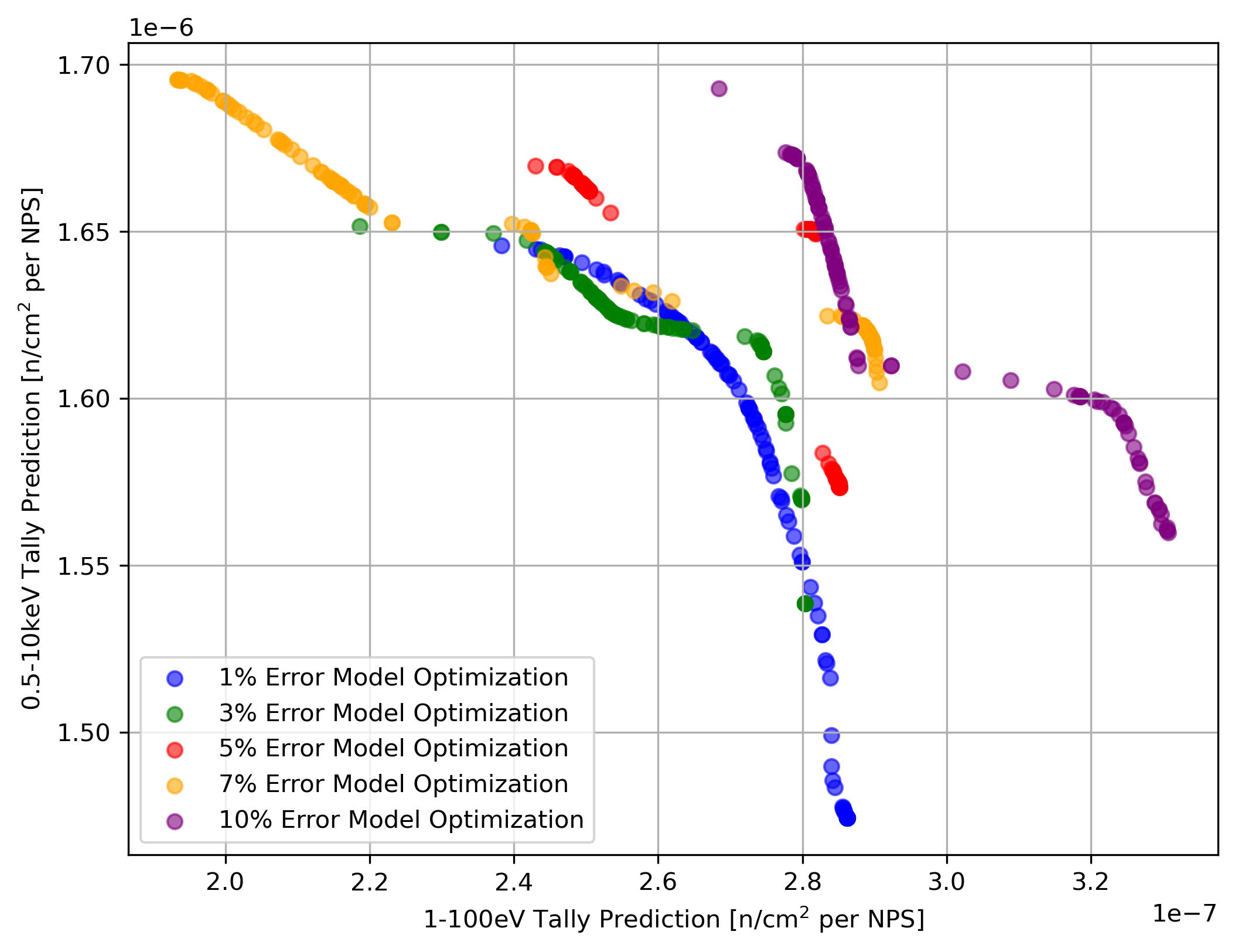}
    \end{subfigure}%
    \hfill
    \begin{subfigure}{0.5\textwidth}
        \centering
        \includegraphics[width=\linewidth]{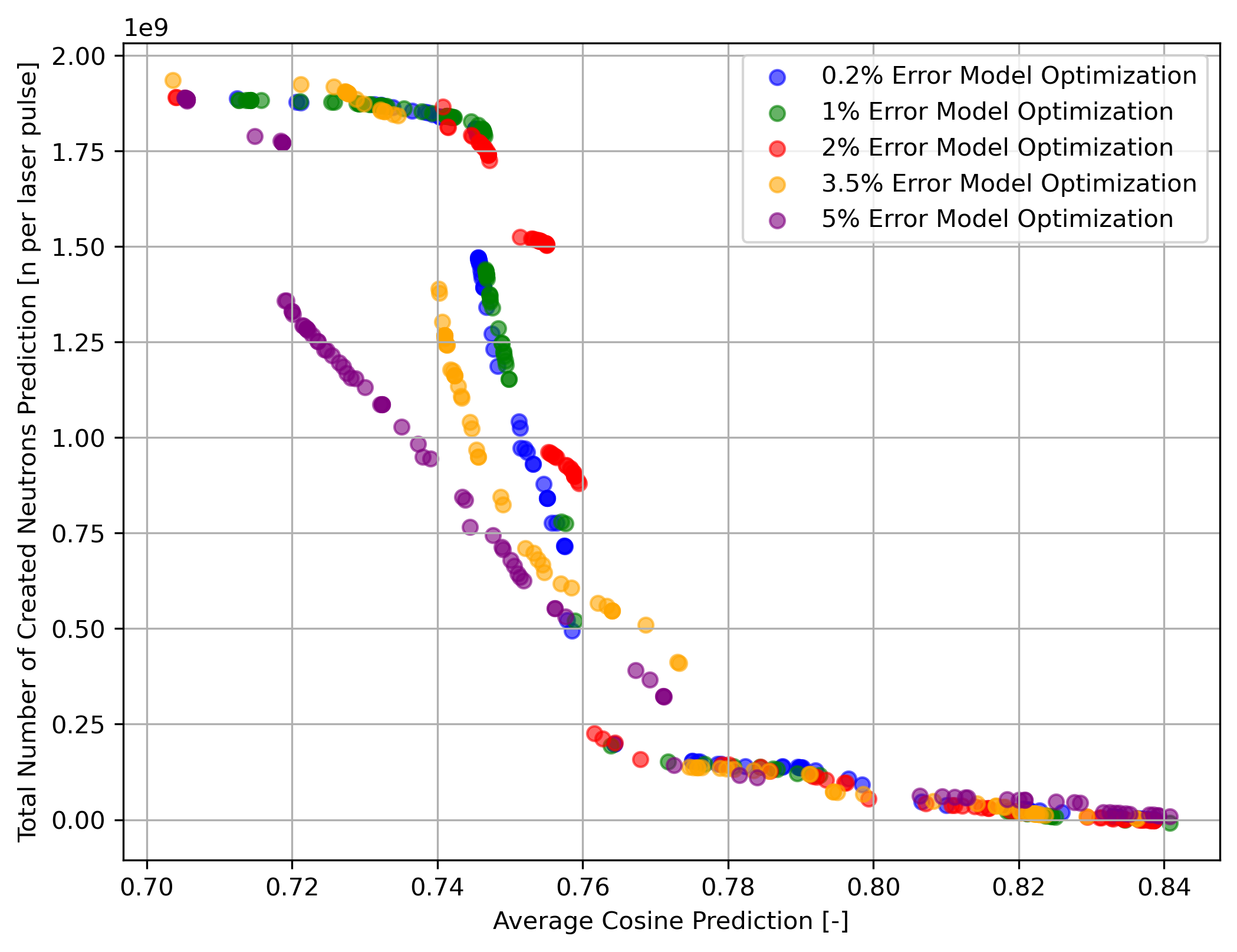}
    \end{subfigure}
    \caption{The optimization Pareto-front curves of the predictions made by using the trained surrogate models and NSGA-III algorithm for neutron moderator design problem (left) and ion-to-neutron converter design problem (right).}
    \label{fig:optimized_points}
\end{figure}

The NSGA-III optimization algorithm successfully generated 100 Pareto-optimal points for each surrogate model, representing a diverse set of candidate solutions on the predicted Pareto-front. These candidate solutions capture the intricate trade-offs inherent in the design problem, as approximated by the surrogate models. However, the predicted pareto-front changes drastically with the tally uncertainty level. Even though the training metrics of the surrogate models are near perfect, the errors carried from the data acquisition present in the predictions made by the models. The effect of these carried errors is the inaccuracy of the trained surrogate models when compared to the actual problem objective function. The change of best point predictions in the optimization results can be seen in Figure \ref{fig:optimized_points}. Please note that some of the points predicted by the surrogate models are not physically achievable in the original problem. This is primarily due to the noise in the training data, which can cause the models to overestimate objective values beyond realistic limits. Optimization results from low-accuracy models that appear to outperform high-accuracy models are often products of these training errors. These inflated predictions reflect the inherent inaccuracies introduced by using data with significant tally uncertainties. 

The NSGA-III algorithm used in the optimization aims to maximize both objectives. However, it is not possible to maximize both objectives at the same time for the selected problems. These two problems present Pareto-fronts which require building an optimized curve with trade-offs of objectives. The Pareto-fronts created by running the optimization algorithm on the created surrogate models are given in Figure \ref{fig:optimized_points}. The best points of this optimization process are located on the Pareto-front and the input parameters corresponding to these Pareto-front optimal points are used in the verification phase using MCNP simulations. In the next phase of our study, the performance of these candidate solutions will be verified through MCNP simulations, enabling us to assess the accuracy and reliability of the surrogate model predictions.

\subsection{Optimization Result Verification}
\label{sec:opt_verif}

For each of the design problems investigated in this work, the neutron moderator design problem and the ion-to-neutron converter design problem, five surrogate models were constructed. The top 100 optimization points from each model were selected, yielding 500 optimized points per problem for the five tally uncertainty levels. The input parameters of these points were then used to generate high-fidelity MCNP simulations with a 0.1\% tally uncertainty cutoff, which will be used as reference simulations. In total, the 1000 points from both problems serve as the verification points for our analysis. Since the NSGA-III optimization algorithm is a continuous optimization algorithm, the points resulting from the optimization have many floating digits. In order to limit the precision of the simulation parameters, the input parameters are rounded to 3 floating points before creating the MCNP simulations of the verification set. 

The value of the normalized hypervolume is used as a comparison metric for the optimization performances. The verification results are normalized in between the maximum and minimum values observed in the verification set simulations. For each objective, the maximum values in the verification results are set to 1, and minimum values are set to 0. The nadir point, combination of the worst observed objectives, was chosen as the reference point for hypervolume calculations. After normalization, this point has been set to (0,0). 

\begin{figure}[!htb]
    \centering
    \begin{subfigure}{0.5\textwidth}
        \centering
        \includegraphics[width=\linewidth]{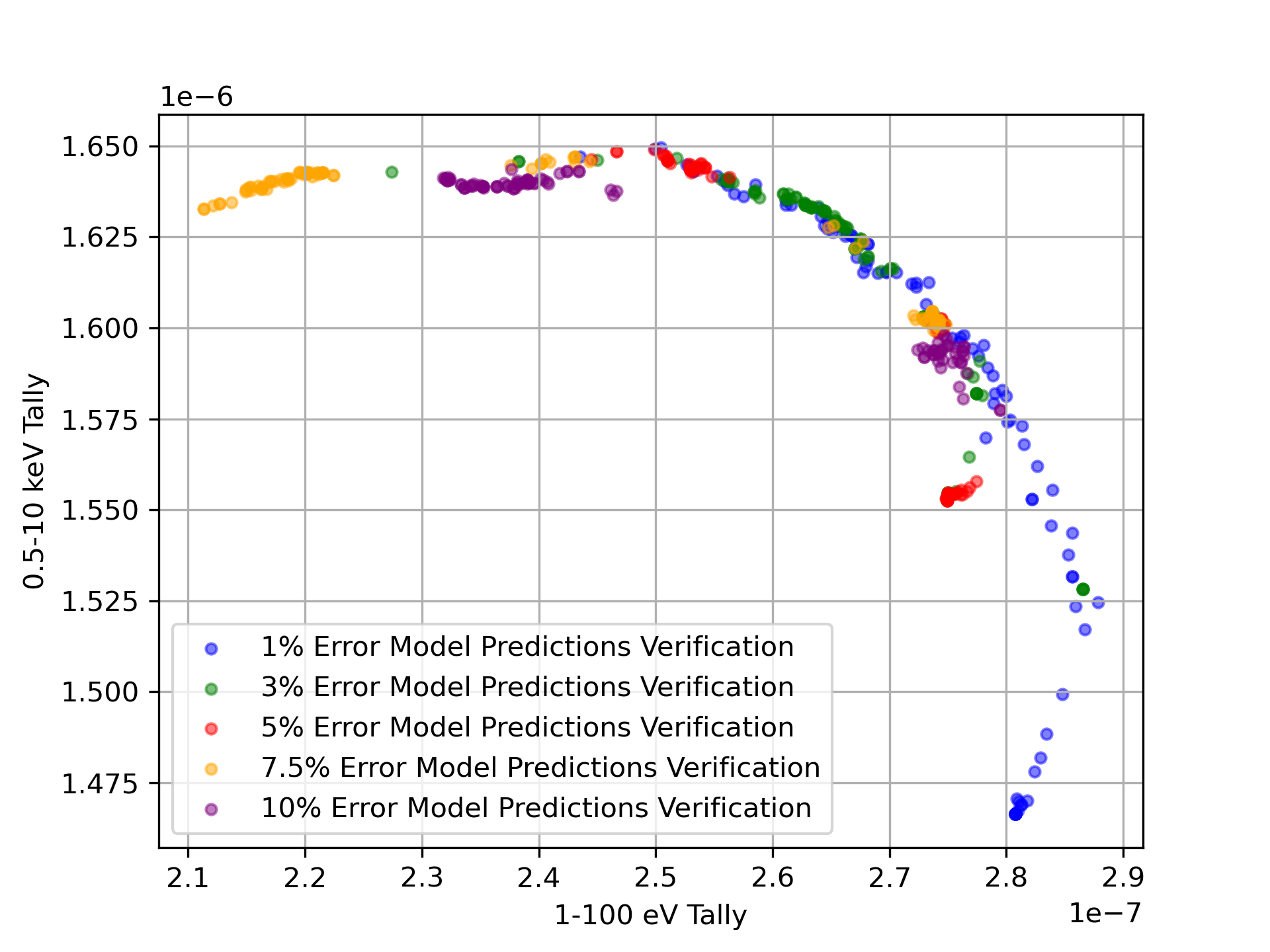}
    \end{subfigure}%
    \hfill
    \begin{subfigure}{0.5\textwidth}
        \centering
        \includegraphics[width=\linewidth]{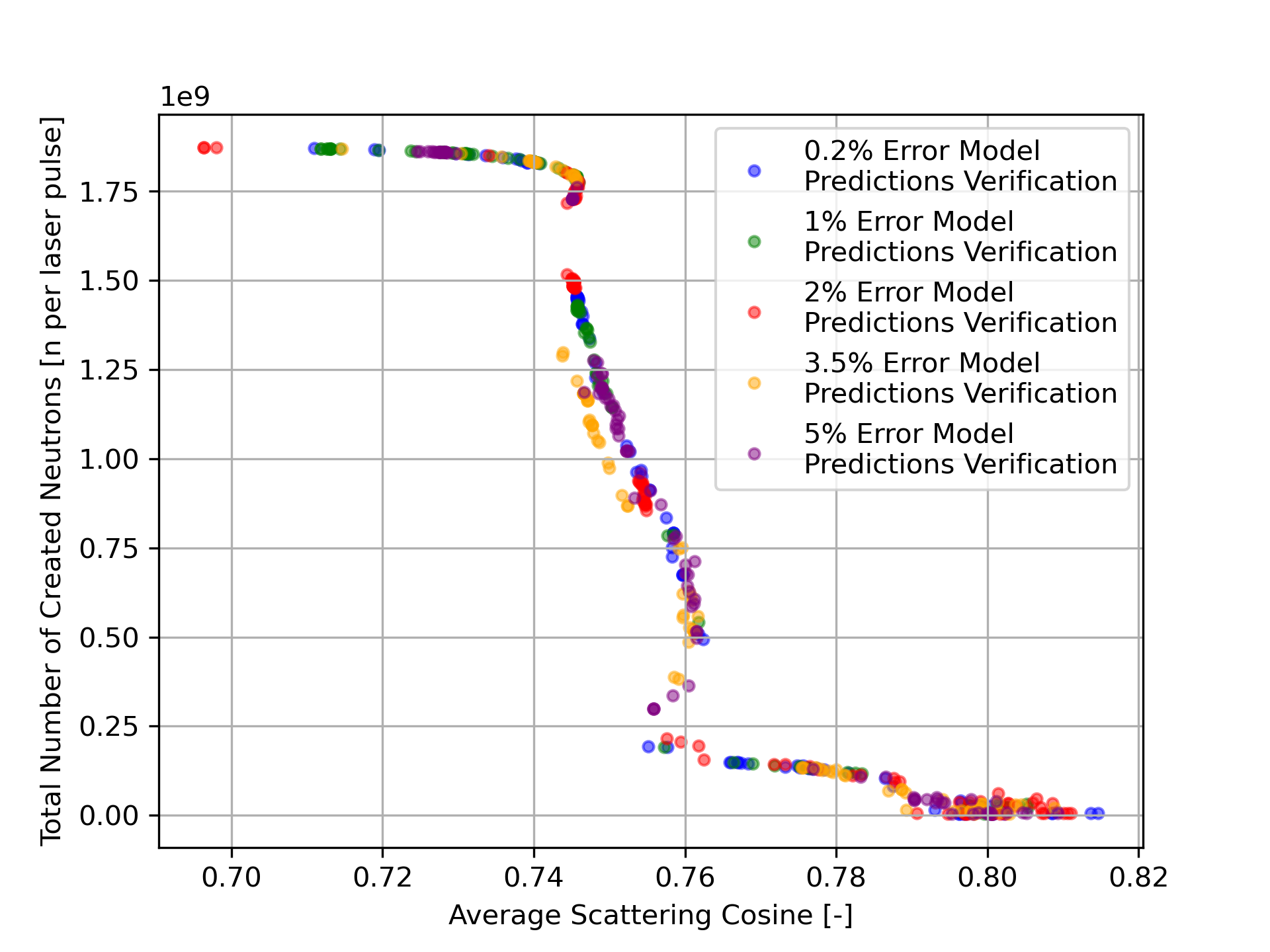}
    \end{subfigure}
    \caption{Verification MCNP simulation results and discovered Pareto-fronts for neutron moderator design problem (left) and ion-to-neutron converter design problem (right).}
    \label{fig:ver_points}
\end{figure}

The impact of surrogate model training data uncertainty on optimization performance reveals contrasting behaviors between the neutron moderator and ion-to-neutron converter design problems. The verification results after the MCNP verification simulations can be seen in Figure \ref{fig:ver_points}. These points are created by the high-fidelity simulations of the exact same input parameters that Figure \ref{fig:optimized_points} used in the surrogate model predictions. In the moderator design problem, surrogate models trained with a 1\% error achieved a normalized hypervolume of 0.8863 relative to the maximum theoretical hypervolume. As the training error was increased, the corresponding hypervolume decreased. In this problem, increasing the tally uncertainty from 1.0\% to 10.0\% results in a monotonic degradation of the normalized hypervolume, with a relative loss exceeding 15\%. This indicates that surrogate model accuracy is critical to maintaining optimization quality in this problem, where higher uncertainties lead to increasingly suboptimal solutions. 

In the ion-to-neutron converter design problem, the relationship between data uncertainty and optimization performance is non-monotonic. While the lowest error (0.2\%) yields the highest normalized hypervolume value of 0.4954, moderate increases in error (e.g., 3.5\% and 5.0\%) still produce near-equivalent performance, with relative losses remaining within 2\%. Notably, the 2.0\% error case results in the lowest hypervolume. However, the reduction is minor, with the resulting value still falling within 5\% of the maximum achievable hypervolume across all tally uncertainty levels considered in this study.

The effect of tally uncertainty in data creation is the same in both problems to the initial data, increasing the spread of the data and a systematic shift. However, the surrogate-models are able to recover the Pareto-front in certain cases, resulting in a complex influence on optimization in this setting. Since the effect of data uncertainty is probabilistic, its data degrading effects may not propagate in every single data uncertainty level. Even though the data quality of the ion-to-neutron converter problem has degraded with the increasing data uncertainty (as shown in Figure \ref{fig:conv_data_plot}), and the obtained Pareto-front has systematically shifted (as shown in Figure \ref{fig:optimized_points}), the surrogate-driven optimization was able to obtain the true Pareto-front unlike the neutron moderator design problem.

From a computational perspective, both problems show substantial speedups with increased uncertainty due to the decreased number of particles used in these simulations. These findings suggest that while certain problems demand high-fidelity surrogates to preserve solution quality, others perform well even under high simulation uncertainty, enabling significant computational savings without compromising optimization performance. 

These findings suggest that the quality of the predicted Pareto front is inherently problem-dependent. In certain problems, increasing the uncertainty in the training data significantly degrades optimization performance, necessitating the use of high-fidelity simulations to preserve solution quality. Conversely, other problems exhibit robustness to surrogate model inaccuracies, where even low-fidelity data (despite its apparent noise) can yield comparable optimization outcomes. This indicates that high simulation accuracy may not always be essential to gain meaningful insights. Identifying the sensitivity of a given problem to training data uncertainty is therefore crucial. Tailoring the fidelity level to the problem's tolerance for error offers a promising strategy to substantially reduce the computational burden of data generation without compromising the effectiveness of the optimization process.

\begin{table}[ht]
    \centering
    \caption{Normalized hypervolume with respect to surrogate model training data tally uncertainty in the neutron moderator design problem.}
    \begin{tabular}{|c|c|c|c|c|c|}
        \hline
        \textbf{Training Data Uncertainty (\%)} & \textbf{1.0} & \textbf{3.0} & \textbf{5.0} & \textbf{7.5} & \textbf{10.0} \\
        \hline
        \textbf{Normalized Hypervolume}  & 0.8863     & 0.8496     & 0.7805     & 0.7640     & 0.7483     \\
        \hline
        \textbf{Relative Hypervolume Loss}  & 0\%     & 4.141\%     & 11.937\%     & 13.190\%     & 15.570\%     \\
        \hline
    \end{tabular}
    \label{tab:normalized_hypervolume_refl}
\end{table}

\begin{table}[ht]
    \centering
    \caption{Normalized hypervolume with respect to surrogate model training data tally uncertainty in the ion-to-neutron converter design problem.}
    \begin{tabular}{|c|c|c|c|c|c|}
        \hline
        \textbf{Training Data Uncertainty (\%)} & \textbf{0.2} & \textbf{1.0} & \textbf{2.0} & \textbf{3.5} & \textbf{5.0} \\
        \hline
        \textbf{Normalized Hypervolume}  & 0.4954     & 0.4892     & 0.4752     & 0.4861     & 0.4899     \\
        \hline
        \textbf{Relative Hypervolume Loss}  & 0\%     & 1.246\%     & 4.772\%    & 1.873\%    & 1.094\%    \\
        \hline
    \end{tabular}
    \label{tab:normalized_hypervolume_conv}
\end{table}

An interesting observation from our study is that some feedforward neural network (FNN) surrogate models trained on low-fidelity simulation data tend to amplify the effects of small input perturbations on the design objectives. As a result, much of the Pareto front predicted by the surrogate trained on the 10\% tally uncertainty data (Figure \ref{fig:optimized_points}) collapses into two narrow regions of the objective space in the verification results (Figure \ref{fig:ver_points}). This collapse is primarily caused by outliers introduced during the noisy data generation process. Due to the stochastic nature of tally estimations in Monte Carlo simulations, high uncertainty can produce data points that deviate significantly from the underlying physical trends. When these outliers are used to train surrogate models, the resulting networks learn to overfit or bias their predictions toward these region. In practice, however, small input variations often do not lead to substantial changes in the design objectives, and the exaggerated predictions seen in the surrogate are largely driven by fluctuations in the MCNP tallies. This leads to a distorted optimization process, where many genuinely optimal solutions are excluded prematurely, causing the Pareto front to shrink as tally uncertainty increases. To mitigate this effect, pre-filtering of the outliers or denoising the training data presents a viable strategy for improving the reliability of surrogate-based optimization, especially when using low-fidelity simulation data.

Lastly, the authors wish to disclose that the analysis in this section was repeated 100 times, beginning with training a new neural network model and continuing through to Pareto front recovery. This was done to assess whether surrogate model training would influence the conclusions. The results from NSGA-III and the hypervolume loss were consistently stable across the 100 runs, indicating that the initial weights and training of the surrogate model did not significantly affect the outcomes.

\section{Concluding Remarks}
\label{sec:conc}

In this study, we investigated the application of feedforward neural network (FNN) surrogate models to complex multi-objective design problems, specifically focusing on neutron physics problems simulated using Monte Carlo tradition transport simulations. The two case studies included neutron moderator design and ion-to-neutron converter design. By leveraging MCNP simulation data at various tally uncertainty levels, we analyzed the impact of surrogate model fidelity (derived from both low-fidelity and high-fidelity simulation results) on multi-objective optimization performance.

Our methodology began with the generation of comprehensive datasets, comprising 3,750 and 8,800 data points for the moderator and converter problems, respectively. These datasets, characterized by varying levels of MCNP tally uncertainty, served as the foundation for training ten distinct surrogate models. Rigorous hyperparameter tuning and performance evaluation using metrics such as MSE and R$^2$ ensured that the surrogate models could replicate simulation outputs accurately, despite the inherent noise in the training data. The trained models were integrated into an NSGA-III multi-objective optimization framework, enabling effective exploration of the design space and identification of Pareto-optimal solutions. With a population size of 100, the optimization converged rapidly (typically within 25 generations) producing a predicted Pareto front for each surrogate model that reflected the trade-offs achievable with its respective fidelity level.

The variance in predicted Pareto fronts was found to be strongly problem-dependent. In the ion-to-neutron converter design problem, low-fidelity simulation data produced optimization results comparable to those achieved using high-fidelity data, suggesting that the impact of training data uncertainty can be decreased by surrogate models. In contrast, the neutron moderator design problem exhibited high sensitivity to data fidelity. In this case, training on low-fidelity data led to exaggerated objective responses in surrogate-models and misleading predictions, many of which could not be verified by accurate simulations or realized in practical designs. These false optima overshadowed the true Pareto front and diminished optimization effectiveness, as evidenced by the decline in normalized hypervolume values. Thus, identifying a problem’s sensitivity to surrogate fidelity is crucial for balancing model accuracy and computational cost in multi-objective optimization.

These observations underscore the critical (but problem-specific) influence of data fidelity on surrogate model performance when conducting design optimization with Monte Carlo simulations. While high-fidelity data generally improves surrogate reliability by reducing noise and limiting sensitivity to small input perturbations, our results demonstrate that this is not always necessary. In some problems, surrogate models that use low-fidelity data can still yield valid and efficient optimizations. Conversely, sensitive problems demand greater fidelity to prevent distortion of the surrogate's predictive behavior. To determine the appropriate fidelity level, preliminary simulations at varying uncertainty levels can be performed and their effects on surrogate behavior and optimization outcomes can be assessed. These findings highlight the potential of \textbf{multi-fidelity simulation strategies} that strike a balance between accuracy and computational efficiency. By aligning fidelity with problem-specific sensitivity, it is possible to achieve reliable optimization results at reduced computational cost.

Overall, the integration of FNN surrogate models with evolutionary optimization techniques offers a promising approach for addressing complex engineering design challenges. Future work should focus on leveraging multi-fidelity data to identify appropriate fidelity requirements, filtering misleading low-fidelity data, and refining surrogate model architectures to enable their compatibility with low-fidelity data. These efforts will further enhance predictive accuracy and the resolution of Pareto fronts when working with high-uncertainty, low-fidelity datasets. The insights gained from this study provide a foundation for developing more robust and cost-effective optimization frameworks for advanced engineering applications.

\section*{Data Availability}
\label{sec:avail}

Currently, the authors possess, in a private GitHub repository, all the data and codes needed to reproduce all the results in this work. To ensure confidentiality of this research, the authors will make this repository public during an advanced stage of the review process, and it will be listed under our research group's public Github page: \url{https://github.com/aims-umich}.

\section*{Acknowledgment}

This work is sponsored by the Department of Energy Office of Nuclear Energy's Distinguished Early Career Program (Award number: DE-NE0009424), which is administered by the Nuclear Energy University Program.  In addition, this work was supported by the U.S. Department of Energy (DOE) through the Los Alamos National Laboratory (LANL). LANL is operated by Triad National Security, LLC, for the National Nuclear Security Administration of the U.S. DOE (Contract No. 89233218CNA000001). 


\section*{CRediT Author Statement}

\begin{itemize}
    \item \textbf{Omer Faruk Erdem}: Methodology, Software, Validation, Formal analysis, Visualization, Data Curation, Investigation, Writing - Original Draft. 
    \item \textbf{Josef Svoboda}: Conceptualization, Methodology, Data Curation, Funding acquisition, Supervision, Resources, Project administration, Writing - Review and Edit.
    \item \textbf{David Paul Broughton}: Conceptualization, Methodology, Data Curation, Funding Acquisition, Supervision, Resources, Project administration, Writing - Review and Edit.
    \item \textbf{Chengkun Huang}: Methodology, Data Curation, Writing - Review and Edit.
    \item \textbf{Majdi I. Radaideh}: Conceptualization, Methodology, Funding acquisition, Supervision, Resources, Project administration, Writing - Review and Edit.
\end{itemize}

\bibliographystyle{elsarticle-num}
\bibliography{references}

\end{document}